\documentclass[12pt]{iopart}

\usepackage{siunitx,graphicx,setstack}

\newcommand{\ket}[1]{\left| #1 \right\rangle}
\newcommand{\bra}[1]{\left\langle #1 \right|}
\newcommand{\inn}[2]{\left\langle #1 \middle| #2 \right\rangle}
\newcommand{\Op}[3]{\left\langle #1 \middle| #2 \middle| #3 \right\rangle}
\newcommand{\avg}[1]{\left\langle #1 \right\rangle}

\newcommand{\Ham}[1]{\hat{H}_{\rm #1}}
\newcommand{\ClHam}[1]{\mathcal{H}_{\rm #1}}
\newcommand{\ZPF}{\mathcal{M}_{\rm ZPF}}
\newcommand{\Texp}[1]{\mathrm{Texp}\left[ #1 \right]}

\makeatletter
\newcommand{\oset}[2]{%
  {\mathop{#2}\limits^{\vbox to 3.1\ex@{\kern-\tw@\ex@
   \hbox{\scriptsize #1}\vss}}}}
\makeatother

\begin{document}

\title{Protocol for generating an arbitrary quantum state of the magnetization in cavity magnonics} 
\author{Sanchar Sharma} 
\address{Max Planck Institute for the Science of Light, 91058 Erlangen, Germany}

\author{Victor A. S. V. Bittencourt} 
\address{Max Planck Institute for the Science of Light, 91058 Erlangen, Germany}

\author{Silvia Viola Kusminskiy}
\address{Institute for Theoretical Solid State Physics, RWTH Aachen University, 52074 Aachen, Germany}
\address{Max Planck Institute for the Science of Light, 91058 Erlangen, Germany}

\begin{abstract}
We propose and numerically evaluate a protocol to generate an arbitrary quantum state of the magnetization in a magnet. The protocol involves repeatedly exciting a frequency-tunable superconducting transmon and transferring the excitations to the magnet via a microwave cavity. To avoid decay, the protocol must be much shorter than magnon lifetime. Speeding up the protocol by simply shortening the pulses leads to non-resonant leakage of excitations to higher levels of the transmon accompanied by higher decoherence. We discuss how to correct for such leakages by applying counter pulses to de-excite these higher levels. In our protocol, states with a maximum magnon occupation of up to $\sim9$ and average magnon number up to $\sim4$ can be generated with fidelity $>0.75$.
\end{abstract}

\maketitle 

Magnets have found commercial applications in magnetic field sensing and storing of information, and are a promising building block for long-range information transfer \cite{Cornelissen15,Kajiwara10} and low-power logic devices \cite{Chumak15}. Recently, there has been an interest in bringing such applications into the quantum domain, known as `quantum magnonics' \cite{Tabuchi_QMag,yuan_quantum_2021,bunkov_quantum_2020,Roadmap}. This is partly fuelled by the extremely low magnetic dissipation found in the ferrimagnetic insulator Yttrium Iron Garnet (YIG) \cite{SagaOfYIG}, along with evidence for macroscopic (mm-long) coherence lengths \cite{Iacopo20,Andrich17}. In the limit of small deviations of spins from their equilibrium value, the collective excitations of a magnet can be modelled as a set of harmonic oscillators called magnons with a lifetime characterized by the Gilbert damping constant of the material that is particularly low for YIG \cite{SagaOfYIG}. Magnons can be probed via their interaction with microwaves (MWs) \cite{SoykalPRL10,Zhang14,TabuchiHybrid14} or optical light \cite{James_OMag,Osada_OMag,Silvia_OMag,Zhang_OMag}. This interaction can be enhanced by employing an electromagnetic cavity to confine the photons, whose decay rate is given by the quality factor of the cavity. The strong coupling regime between cavity MW photons and magnons, where information can be coherently exchanged between the two excitations at rates $>\SI{10}{\mega\hertz}$ much faster than each of their decay rates $<\SI{1}{\mega\hertz}$, can be routinely achieved in experiments. The strong coupling regime in the optical domain is notoriously more difficult to achieve, however there are promising developments in this direction \cite{James_TinyCavity,Jasmin_Vortex,Jasmin_Crystal,Victor_ENZ,Zhu_MO_WG,ModeMatching}. 

A quantum platform involving magnets requires generation, manipulation, and detection of non-classical states of the magnetization. Possible applications in the quantum regime include magnon-based quantum transducers between light and MWs, as well as transduction involving other degrees of freedom such as phonons \cite{Zhang16,Carlos_MagPh} and electrons \cite{Heinrich_SPCoupling,Li_SPCoherence,Woltersdorf_SPBackflow}. Furthermore, non-classical magnetization states can be useful in magnetic field sensing reaching the Heisenberg limit \cite{BeyondSQL}. While single-magnon detection has been demonstrated \cite{MagDet1,MagDet2,MagDet3} by exploiting the coupling of magnets to superconducting circuits mediated by MW cavities \cite{Tabuchi_QMag,TabuchiQubit15,Lachance_Hybrid,Babak_CavityMagnonics}, experiments so far have probed coherent or thermal magnon states whereas ``true'' non-classical states (defined e.g. by a Wigner function with negative regions) remain to be realized. Theoretical proposals in this direction include the generation of cat states of the magnetization using MW cavity photons, \cite{CatStateMW} and the generation of single magnon Fock states \cite{Victor_Heralding} and cat states \cite{CatStateOptical} via optical means. In thin films, the generation of entangled pairs of traveling magnons was proposed \cite{Mehrdad_NCMags}. All of these proposals are probabilistic for the state generation (except for the last) and are specialized to their respective target states, while there is no known method to deterministically generate an arbitrary magnon state. 

\begin{figure}
	\[ \includegraphics[width=0.7\columnwidth]{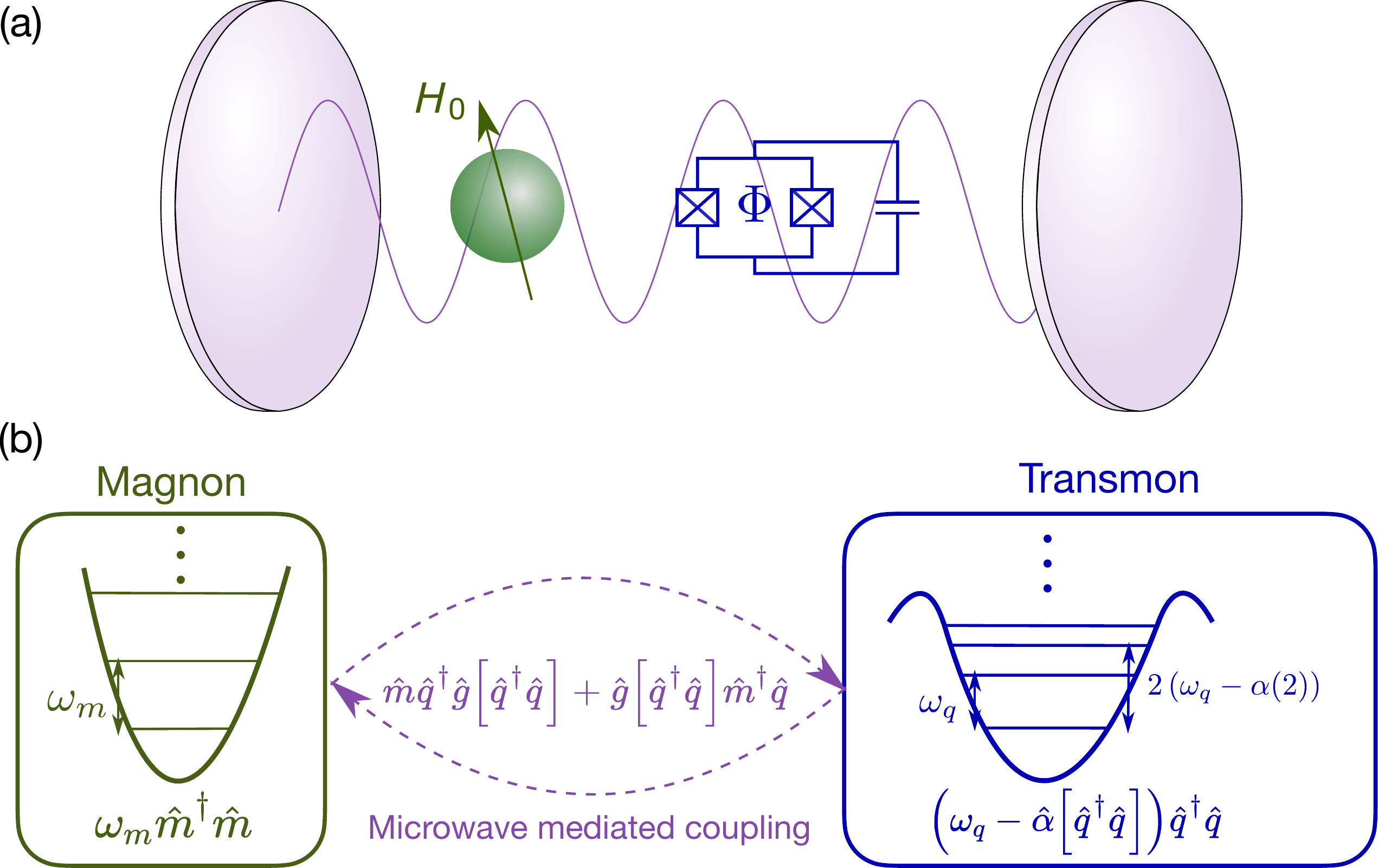} \]
	\caption{(a) The setup consists of a MW cavity loaded with a magnet and a superconducting transmon. The frequency of the Kittel mode can be tuned by an external magnetic field, while that of the transmon by using an input flux $\Phi$. (b) Model of the system as a harmonic oscillator (magnons) coupled to an anharmonic oscillator (transmon).  The coupling rate between magnon and transmon via the MWs depends on the transmon occupation. All the terms in the figure are defined in Sec. \ref{sec:System}.} \label{Fig:Schematic}
\end{figure}

In this manuscript we fill in this gap by proposing a protocol to this end, tailored for magnets coupled to a superconducting transmon via a MW cavity, as depicted in Fig. \ref{Fig:Schematic}. Our method is based on the protocol proposed in Ref. \cite{Law96}, which shows that the deterministic generation of an arbitrary quantum state of a harmonic oscillator can be accomplished by coupling the oscillator to a qubit. A superconducting transmon is a weakly anharmonic oscillator whose first two levels can be treated as a qubit. We show how the protocol of Ref. \cite{Law96} can be modified to take into account the anharmonicity of the superconducting transmon, in order to generate arbitrary quantum magnon states with high fidelity. 

The paper is organized as follows. We discuss the system and effective Hamiltonian in Sec. \ref{sec:System}. In Sec. \ref{sec:Gen:TLS} we approximate the superconducting transmon by a two-level system and review the protocol from Ref. \cite{Law96} for the cavity mangnonic system. In Sec. \ref{sec:Gen:Transmon} we discuss how to correct for errors due to deviations from the two-level approximation, in particular due to interferences from higher energy levels present in the transmon. In Sec. \ref{sec:Results}, we discuss the fidelity of state generation for paradigmatic examples of non-classical states. We discuss possible improvements to our protocol that can increase the fidelity of generation further in Sec. \ref{sec:Improvements}. We review the main findings qualitatively in Sec. \ref{sec:Conclusions}. 

\section{System\label{sec:System}} 
 
The hybrid system we considered is depicted in Fig. \ref{Fig:Schematic}: inside a MW cavity, magnons can couple to a superconducting transmon \cite{Tabuchi_QMag,TabuchiQubit15}. The effective coupling Hamiltonian can be obtained by tracing out the MW field, as we review in this section. 

When the magnet is much smaller than the wavelength of the MWs ($\sim\si{\centi\meter}$), only the total magnetization $\bi{M}$ couples to the MWs, so we need to consider only the uniform magnon mode known as the Kittel mode. It can be quantized via a Holstein-Primakoff transformation \cite{HolPrim,StanPrabh} (see \ref{app:Ham_eff} for details) that, in the limit of $M_{x,y} \ll M_z$, reduces to
\begin{equation}
	 M_x - iM_y\rightarrow2\ZPF\hat{M},\ M_z\rightarrow M_s - \frac{2\ZPF^2}{M_s}\hat{M}^{\dagger}\hat{M} . \label{Def:AnnOp} 
\end{equation} 
where $\hat{M}$ is the magnon annihilation operator satisfying the bosonic commutation relations $[\hat{M},\hat{M}^{\dagger}] = 1$, $M_s$ is the saturation magnetization, and the zero-point fluctuations of the magnetization are given by 
\begin{equation}
	 \ZPF = \sqrt{\frac{\gamma\hbar M_s}{2V_m}},\label{Def:ZPF} 
\end{equation} 
with $\gamma>0$ being the absolute gyromagnetic ratio and $V_m$ being the volume of the magnet. Assuming a spherical magnet and an external magnetic field $H_{{\rm app}}$, the Hamiltonian becomes $\Ham{m} = \hbar\Omega_m\hat{M}^{\dagger}\hat{M}$ with $\Omega_m = \gamma\mu_0H_{{\rm app}}$ (see \ref{app:Ham_eff}) which can be tuned to be in the range of $\sim2\pi\times5\text{-}\SI{10}{\giga\hertz}$. For a YIG sample with volume $V_m>\SI{1}{\micro\meter\cubed}$, $M_s = \SI{140}{\kilo\ampere\per\meter}$, and $\gamma = 2\pi\times\SI{28}{\giga\hertz\per\tesla}$, we have $\ZPF/M_s<8\times10^{- 6}$, implying that the deviations from the ground state are small. 

A flux-tunable transmon consists of two Josephson junctions in parallel with Josephson energy $E_{J1,J2}$ respectively, forming a SQUID loop along with a capacitor with capacitance $C$ also in parallel \cite{Blais_CQED,Blais_RMP,Koch_Transmon,Krantz_SCQubits}. When the charging energy $E_C = e^2/2C$ is such that $E_C\ll E_{J1,J2}$ , the energy levels of the circuit resemble that of a weakly anharmonic oscillator, 
\begin{equation}
	 E_n \approx \hbar\left(\Omega_q - \frac{A}{2}(n - 1)\right)n, 
\end{equation} 
where $A = E_C/4$ and $\Omega_q = \sqrt{E_CE_{J,{\rm eff}}}$ with the effective Josephson energy $E_{J,{\rm eff}}$ being 
\begin{equation}
	 E_{J,{\rm eff}}(\Phi) = \sqrt{\left(E_{J1} + E_{J2}\right)^2\cos^2\frac{\pi\Phi}{\Phi_0} + \left(E_{J1} - E_{J2}\right)^2\sin^2\frac{\pi\Phi}{\Phi_0}} . 
\end{equation} 
Here $\Phi$ is the flux inside the SQUID loop and $\Phi_0 = h/2e$ is the superconducting flux quantum. Such a set of energy levels can be labelled using an annihilation operator $\hat{Q}$ and its corresponding number operator $\hat{Q}^{\dagger}\hat{Q}$. The number of quanta exchanged between the two Josephson junctions is given by 
\begin{equation}
	 \hat{N}_{{\rm sc}} = \left(\frac{E_{J,{\rm eff}}}{2E_C}\right)^{1/4}\frac{\hat{Q} - \hat{Q}^{\dagger}}{\sqrt{2}i} . 
\end{equation} 
The Hamiltonian becomes 
\begin{equation}
	 \Ham{tr} = \hbar\left[ \Omega_q - \frac{\hbar A}{2}\left(\hat{Q}^{\dagger}\hat{Q} - 1\right) \right] \hat{Q}^{\dagger}\hat{Q}. 
\end{equation} 
The frequency $\Omega_q$ can be tuned dynamically via the external flux in SQUID $\Phi$, to be in a typical range of $2\pi\times5\text{-}\SI{20}{\giga\hertz}$. Tuning via $\Phi$ can be achieved at a fast timescale $\ll\SI{10}{\nano\second}$ \cite{Rol19,Hofheinz09}, but it comes at the cost of increased decoherence proportional to the allowed tunability $\left|E_{J1} - E_{J2}\right|$, as a higher tunability implies larger susceptibility to flux noises \cite{Hutchings_17,Koch_Transmon}. For our case, we assume that $\Omega_q$ is tunable within a range of $\sim2\pi\times\SI{1}{\giga\hertz}$. 

We assume that the magnet and the transmon are both placed inside a MW cavity which can be modelled as a set of harmonic modes with frequencies $\Omega_{a,r}$ and annihilation operator $\hat{A}_r$. The details of the modes depend on the shape and size of the cavity. The Hamiltonian for the unloaded cavity is $\Ham{mw} = \sum_r\hbar\Omega_{a,r}\hat{A}_r^{\dagger}\hat{A}_r$. Its coupling to both the magnet and the transmon corresponds to exchange of quanta and can be modelled as [see \ref{app:Ham_eff}] 
\begin{equation}
	 \frac{\Ham{int}}{\hbar} = \sum_r\left(G_{m,r}\hat{M}^{\dagger}\hat{A}_r + G_{m,r}^*\hat{M}\hat{A}_r^{\dagger} + G_{q,r}\hat{Q}^{\dagger}\hat{A}_r + G_{q,r}^*\hat{Q}\hat{A}_r^{\dagger}\right), 
\end{equation} 
where the coupling $G_{m,r}\propto\left|\bi{B}(\bi{r}_{{\rm magnet}})\right|$ with $\bi{B}(\bi{r}_{{\rm magnet}})$ the magnetic field at the location of the magnet, and $G_{q,r}\propto\left|\bi{E}(\bi{r}_{{\rm transmon}})\right|$ where $\bi{E}(\bi{r}_{{\rm transmon}})$ is the electric field at the location of the transmon. If for all $r$ modes, $\Omega_{a,r} - \Omega_m>G_{m,r}$ and $\Omega_{a,r} - \Omega_q>G_{q,r}$, we can define cavity-dressed magnon and transmon modes with annihilation operators $\hat{m}$ and $\hat{q}$ respectively. Up to quadratic order in the couplings \cite{Tabuchi_QMag}, 
\begin{equation}
	 \frac{\hat{H}}{\hbar} = \omega_m\hat{m}^{\dagger}\hat{m} + \left(\omega_q - \hat{\alpha}\left[\hat{q}^{\dagger}\hat{q}\right]\right)\hat{q}^{\dagger}\hat{q} + \hat{m}\hat{q}^{\dagger}\hat{g}\left[\hat{q}^{\dagger}\hat{q}\right] + \hat{g}\left[\hat{q}^{\dagger}\hat{q}\right]\hat{m}^{\dagger}\hat{q} . \label{eq:Ham_red} 
\end{equation} 
Details about this Hamiltonian are provided in \ref{app:Ham_eff}, while here we list down the features relevant for the rest of the article. The renormalized magnon's frequency $\omega_m$ and transmon's frequency $\omega_q$ stay close to their bare values $\Omega_m$ and $\Omega_q$ respectively. A schematic representation of each term of the Hamiltonian is shown in Fig. \ref{Fig:Schematic}. 

The function $\hat{\alpha}$ gives the anharmonicity of the transmon and $\hat{g}$ gives the magnon-transmon coupling. Below, we use the same characters $\alpha(n) = \Op{n}{\hat{\alpha}}{n}$ and $g(n) = \Op{n}{\hat{g}}{n}$ for the values of anharmonicity and coupling. Typically $\alpha(2)\sim2\pi\times150\text{-}\SI{300}{\mega\hertz}$ increasing with $n$ and the coupling $g$ can achieve $\sim2\pi\times20\text{-}\SI{30}{\mega\hertz}$ depending on the system parameters as discussed in \ref{app:Ham_eff}.  

Restricting the analysis to only the first two levels of the transmon, we get the Jaynes-Cummings Hamiltonian 
\begin{equation}
	 \frac{\hat{H}^{(2)}}{\hbar} = \omega_m\hat{m}^{\dagger}\hat{m} + \omega_q\ket{e}\bra{e} + g\left(\hat{m}\hat{\sigma}_+ + \hat{m}^{\dagger}\hat{\sigma}_-\right),\label{eq:JCHamiltonian} 
\end{equation} 
where $g\equiv g(0)$, $\{\ket{g},\ket{e}\}$ are the first two excited states of the transmon, $\hat{\sigma}_+ = \ket{g}\bra{e}$ and $\hat{\sigma}_- = \ket{e}\bra{g}$. As mentioned before, the frequency $\omega_q$ is tunable. When $\left|\omega_q - \omega_m\right|\gg g$, the magnons and transmon are decoupled besides a negligible dispersive coupling, while at resonance, $\omega_q = \omega_m$, the two subsystems can exchange quanta. 

\section{Ideal State Generation\label{sec:Gen:TLS}} 

\begin{figure}
	\[ \includegraphics[width=0.7\columnwidth]{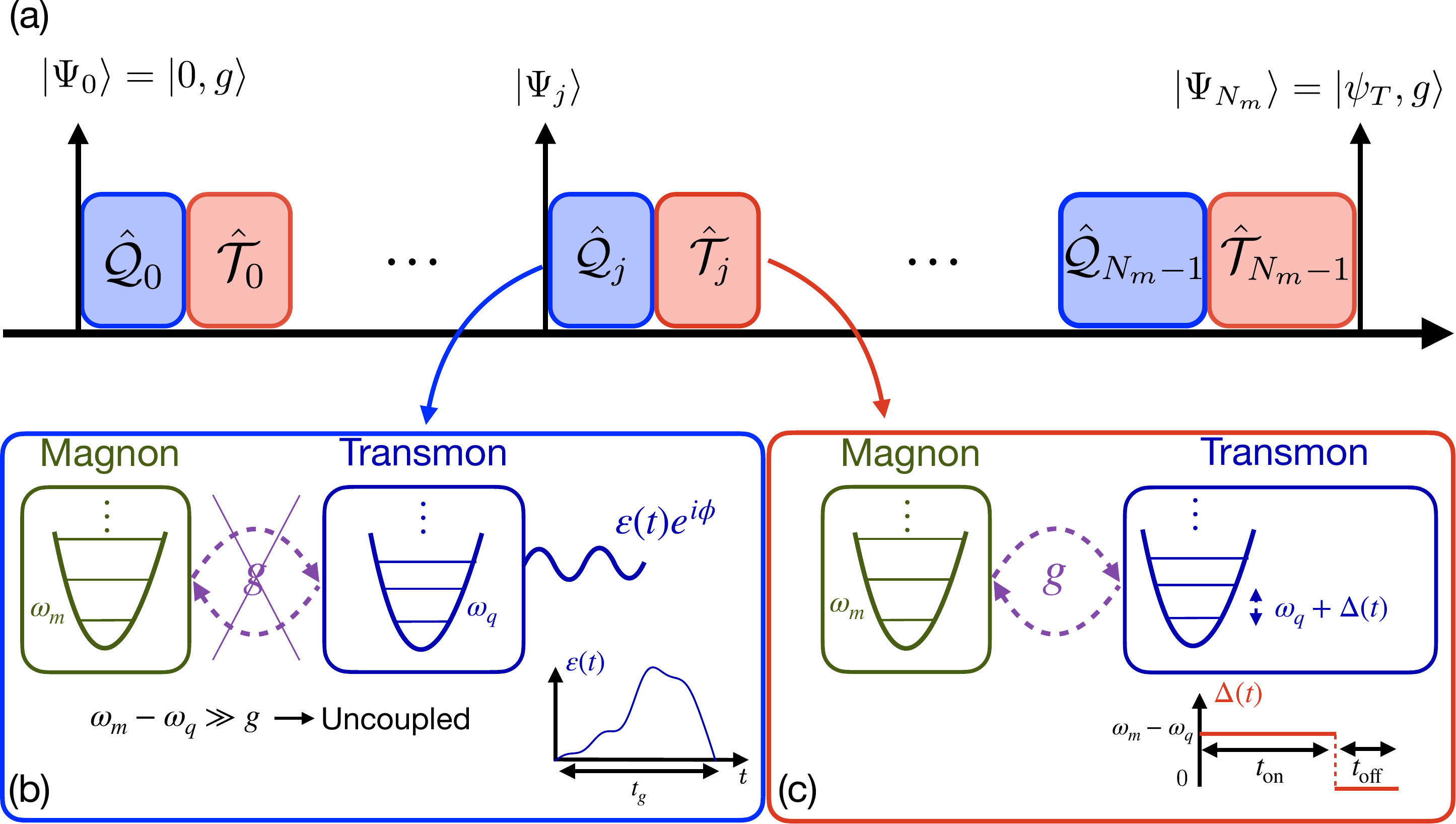} \]
	\caption{(a) Schematic representation of the protocol for generation of an arbitrary magnon state, starting from $\ket{0,g}$, i.e. both magnon and transmon being in ground state, ending in $\ket{\psi_{T},g}$, i.e. the magnon being in a target state $\ket{\psi_{T}}$ and transmon in ground state. (b) The qubit gate is implemented via a time-dependent drive $\tilde{\varepsilon}(t)=\varepsilon(t)e^{i\phi}e^{-i\omega_{q}t}$, and (c) the quanta exchange is accomplished by tuning the transmon's frequency via the term $\Delta(t)$, see Eq. \ref{eq:timedependentterms}.  The target superposition is constructed by choosing appropriate $\{\varepsilon(t),\phi\}$ and $\{t_{{\rm on}},t_{{\rm off}}\}$ at each stage of the protocol.} \label{Fig:ProtocolOper}
\end{figure}

In this section we review the protocol to generate an arbitrary state of the magnetization via its coupling to a two-level system. This is discussed in \cite{Law96} theoretically for a general harmonic oscillator. The protocol consists of: (1) exciting the qubit; (2) partially transfering the excitation to the oscillator and (3) repeating the process multiple times to construct the Fock-state superposition that describes the target state. In this setup, the transmon can be excited via a time-dependent drive, while the transmon and the magnons can be put in and out of resonance by tuning the magnon frequency (via an external magnetic field) or the transmon frequency (via an applied flux), effectively changing the coupling rate (see Fig. \ref{Fig:ProtocolOper}). The procedure was experimentally implemented to generate quantum states of MW photons coupled to a superconducting transmon \cite{Hofheinz09}. 

If we start from the qubit's state $c_0\ket{e} + c_1\ket{g}$ while the magnon is in the ground state $\ket{0}$, switching the coupling on for a period $t = \pi/(2g)$ in the absence of dissipation evolves the state as 
\begin{equation}
	 \left(c_0\ket{e} + c_1\ket{g}\right)\ket{0}\rightarrow\ket{g}\left(c_0\ket{0} + c_1\ket{1}\right), 
\end{equation} 
where $\ket{n}$ is $n$-magnon state. Thus, we can create an arbitrary superposition of $\ket{0}$ and $\ket{1}$. If we repeat this process $N$ times, we can, in principle, generate an arbitrary magnon state containing up to a maximum of $N$ magnons. The protocol then consists of a series of operations in which first the transmon is excited, and then the qubit's frequency is changed to bring it in resonance with the magnons for a time of the order $\pi/g$. 

To make the process concrete, consider the Hamiltonian restricted to the first two levels of the transmon $\hat{H}^{(2)}(t) = \Ham{s}^{(2)} + \Ham{d}^{(2)}(t)$, where the static part is 
\begin{equation}
	 \frac{\Ham{s}^{(2)}}{\hbar} = \omega_m\hat{m}^{\dagger}\hat{m} + \omega_q\ket{e}\bra{e} + g\left(\hat{m}\hat{\sigma}_+ + \hat{m}^{\dagger}\hat{\sigma}_-\right), 
\end{equation} 
and the dynamic part is 
\begin{equation}
	 \frac{\Ham{d}^{(2)}}{\hbar} = \Delta(t)\ket{e}\bra{e} + \tilde{\varepsilon}(t)\hat{\sigma}_+ + \tilde{\varepsilon}^*(t)\hat{\sigma}_-,\label{eq:timedependentterms} 
\end{equation} 
Here and in what follows, the superscript (2) indicates the approximation of the transmon as a two-level system. In this expression, $\Delta(t)$ is the externally induced change in the transmon's frequency, and $\tilde{\varepsilon}(t)$ is the transmon drive amplitude. We assume that the qubit and the magnons are far detuned, $\left|\omega_m - \omega_q\right|\gg g$, implying that magnons and qubit are decoupled at $\Delta = 0$.  

A non-zero $\tilde{\varepsilon}$ can excite a qubit. We define the time evolution operator $\hat{\mathcal{Q}}^{(2)}$ for $\Delta = 0$ and $\tilde{\varepsilon}(t) = \varepsilon(t)e^{- i\omega_qt + i\phi}$, 
\begin{equation}
	 \hat{\mathcal{Q}}^{(2)}\left(\theta,\phi\right) = \Texp{- \frac{i}{\hbar}\int_0^{t_g}dt\left(\Ham{s}^{(2)} - \hbar\varepsilon(t)e^{- i\omega_qt + i\phi}\hat{\sigma}_+ +{\rm h . c .}\right)},\label{Def:Gate:TLS} 
\end{equation} 
where $t_g$ is a pre-determined gate time, $\varepsilon(t)$ is chosen satisfying $\int_0^{t_g}d\tau\varepsilon(\tau) = \theta$ and ${\rm Texp}$ is the time-ordered exponential. As derived in \ref{app:Gates}, 
\begin{equation}
	 \hat{\mathcal{Q}}^{(2)}\left(\theta,\phi\right) = e^{ - i\omega_mt_g\hat{m}^{\dagger}\hat{m} - i\omega_qt_g\ket{e}\bra{e}} \left(\hat{I}\cos\theta + i\sin\theta\left(e^{i\phi}\hat{\sigma}_+ + e^{- i\phi}\hat{\sigma}_-\right)\right), 
\end{equation} 
which applies a Bloch rotation of magnitude $2\theta$ around the axis $(\cos\phi, - \sin\phi,0)$. After the qubit is excited, we want to (partially) transfer the quanta to magnons. We then define the time evolution operator for a detuning $\Delta_0 = \omega_m - \omega_q$ and no qubit driving 
\begin{equation}
	 \hat{\mathcal{T}}^{(2)}\left(t_{{\rm on}},t_{{\rm off}}\right) = \exp\left[\frac{- i\Ham{s}^{(2)}t_{{\rm off}}}{\hbar}\right]\exp\left[ - \frac{i}{\hbar}\left(\Ham{s}^{(2)} + \Delta_0\ket{e}\bra{e}\right)t_{{\rm on}}\right] . \label{Def:Trans:TLS} 
\end{equation} 
An additional `off' time, $t_{{\rm off}}$, adjusts the relative phases in the wave-function \footnote{In \cite{Law96}, the coupling was assumed to be a freely varying complex parameter, so the phases could be adjusted via the phase of $g$.}. As derived in \ref{app:Transfer}, $\hat{\mathcal{T}}^{(2)} = \hat{\mathcal{T}}_{{\rm ph}}^{(2)}\hat{\mathcal{T}}_{{\rm ex}}^{(2)}$ where the phases evolve as 
\begin{equation}
	 \hat{\mathcal{T}}_{{\rm ph}}^{(2)} = \exp\left( - i\omega_m\left(t_{{\rm off}} + t_{{\rm on}}\right)\hat{m}^{\dagger}\hat{m} - i\left(\omega_mt_{{\rm on}} + \omega_qt_{{\rm off}}\right)\ket{e}\bra{e}\right), 
\end{equation} 
and the quanta are exchanged via 
\begin{equation}
	 \hat{\mathcal{T}}_{{\rm ex}}^{(2)} = \cos\left(gt_{{\rm on}}\sqrt{\hat{N}}\right) - \frac{i}{\sqrt{\hat{N}}}\sin\left(gt_{{\rm on}}\sqrt{\hat{N}}\right)\left(\hat{m}\hat{\sigma}_+ + \hat{m}^{\dagger}\hat{\sigma}_-\right), 
\end{equation} 
with the total number operator $\hat{N} = \hat{m}^{\dagger}\hat{m} + \ket{e}\bra{e}$. 

A given target state $\ket{\psi_T}$ containing a maximum of $N_m$ magnons, can be systematically achieved by adding magnons one by one as described above. For that, we need a set of time evolution operators $\hat{\mathcal{Q}}_j^{(2)}\equiv\hat{\mathcal{Q}}^{(2)}(\theta_j,\phi_j)$ and $\hat{\mathcal{T}}_j^{(2)}\equiv\hat{\mathcal{T}}^{(2)}(t_{{\rm on},j},t_{{\rm off},j})$ (see Fig. \ref{Fig:ProtocolOper}) such that 
\begin{equation}
	 \ket{\Psi_0} = \ket{0,g},\ \ket{\Psi_{j + 1}} = \hat{\mathcal{T}}_j^{(2)}\hat{\mathcal{Q}}_j^{(2)}\ket{\Psi_j},\ \ket{\Psi_{N_m}} = \ket{\psi_T,g}, 
\end{equation} 
and each $\ket{\Psi_j}$ has a maximum of $j$ excitations. Explicitly, for $\ket{\Psi_j} = \ket{\psi_j^e,e} + \ket{\psi_j^g,g}$ there are a maximum of $(j - 1)$ and $j$ magnons in $\ket{\psi_j^e}$ and $\ket{\psi_j^g}$ respectively.

\begin{figure}
	\[ \includegraphics[width=0.7\columnwidth]{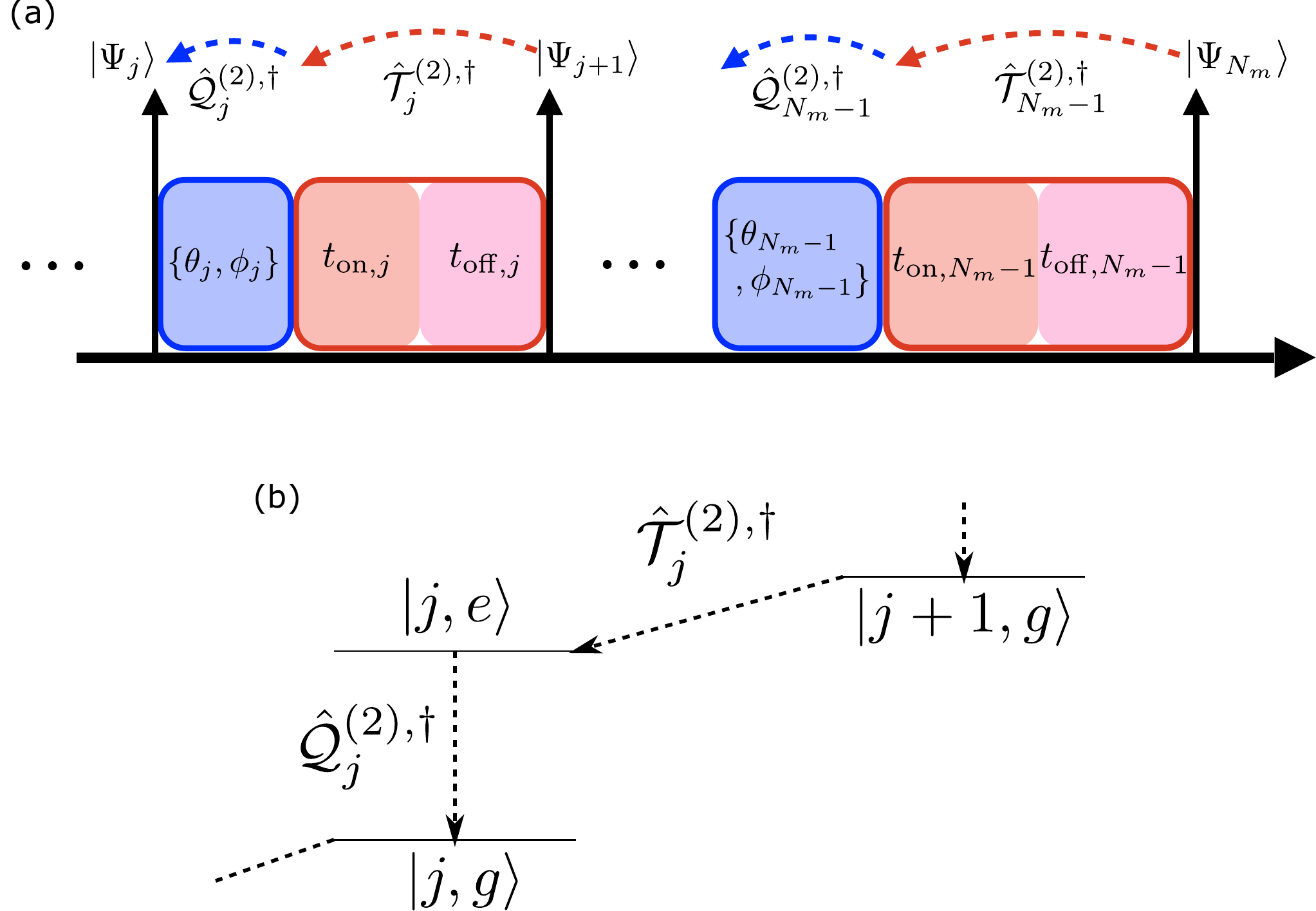} \]
	\caption{(a) The parameters are determined inductively from the last operation to the first. Starting from the target state $\vert\Psi_{N_{m}}\rangle=\vert\psi_{T},g\rangle$, each application of $\left(\hat{\mathcal{T}}^{(2)}\hat{\mathcal{Q}}^{(2)}\right)^{\dagger}$ removes one quanta from the magnon-transmon system. (b) $\hat{{\cal T}}_{j}^{(2),\dagger}$ is chosen to ensure $\Op{j+1,g}{\hat{{\cal T}}_{j}^{(2),\dagger}}{\Psi_{j+1}}=0$ that can be interpreted as transferring amplitude from $\ket{j+1,g}$ to $\ket{j,e}$. Similarly, ensuring $\Op{j,e}{\hat{{\cal Q}}_{j}^{(2),\dagger}\hat{{\cal T}}_{j}^{(2),\dagger}}{\Psi_{j+1}}=0$ can be interpreted as transferring the amplitude from $\ket{j,e}$ to $\ket{j,g}$. \label{fig:TLSProtocol}}
\end{figure}

The coefficients $\{\theta_j,\phi_j,t_{{\rm off},j},t_{{\rm on},j}\}$ are found by reversing the problem in order to systematically remove magnons one by one from $\ket{\Psi_{N_m}}$. Inductively, we assume that $\ket{\Psi_{j + 1}}$ is found. Explicitly, we find $\{t_{{\rm on},j},t_{{\rm off},j}\}$ to ensure $\Op{j + 1,g}{\hat{{\cal T}}_j^{(2),\dagger}}{\Psi_{j + 1}} = 0$ achieved via [see \ref{app:Transfer}] 
\begin{equation}
	 \fl \inn{j + 1,g}{\Psi_{j + 1}}\cos\left(gt_{{\rm on},j}\sqrt{j + 1}\right) + i\inn{j,e}{\Psi_{j + 1}}\sin\left(gt_{{\rm on},j}\sqrt{j + 1}\right)e^{- i\Delta_0t_{{\rm off},j}} = 0 . \label{eq:Trans_res:TLS} 
\end{equation} 
Next, we find $\{\theta_j,\phi_j\}$ to ensure $\Op{j,e}{\hat{{\cal Q}}_j^{(2),\dagger}\hat{{\cal T}}_j^{(2),\dagger}}{\Psi_{j + 1}} = 0$ giving [see \ref{app:Gates}] 
\begin{equation}
	 \Op{j,e}{\hat{{\cal T}}_j^{(2),\dagger}}{\Psi_{j + 1}}\cos\theta_j + i\Op{j,g}{\hat{{\cal T}}_j^{(2),\dagger}}{\Psi_{j + 1}}\sin\theta_je^{i\phi_j - i\omega_qt_g} = 0 . \label{eq:Gate_res:TLS} 
\end{equation} 
Then, we find $\ket{\Psi_j} = \left(\hat{\mathcal{T}}_j^{(2)}\hat{\mathcal{Q}}_j^{(2)}\right)^{\dagger}\ket{\Psi_{j + 1}}$ and the induction continues until all quanta are removed. The procedure to inductively obtain the protocol's parameters is schematically shown in Fig. \ref{fig:TLSProtocol}. 

\section{Realistic State Generation\label{sec:Gen:Transmon}} 

The protocol discussed in the previous section is suitable to generate states with only a small number of excitations. States with higher number occupation require longer protocols and hence are spoiled by dissipation. The protocol duration can be decreased via two ways. Firstly, the gate time to excite the qubit can be decreased. Nevertheless, for a transmon qubit, a shorter gate time, $t_g\sim1/\alpha(n)$ where $\alpha$ is the level-dependent anharmonicity [see below Eq. (\ref{eq:Ham_red})], implies a broader frequency spectrum allowing for significant non-resonant transitions to higher transmon levels. Secondly, a stronger coupling between magnons and the transmon speeds up the excitation transfer step. However, a very strong coupling, $g(n)\sqrt{n}\sim\alpha(n)$ where $g$ is the level-dependent magnon-transmon coupling [see below Eq. (\ref{eq:Ham_red})], implies a leakage of magnon excitations into higher transmon states. Such interferences limit the protocol time and eventually the largest size of the harmonic oscillator state that can be created. In this section, we show that transitions to higher levels can be incorporated by adding extra pulses for canceling out such transition amplitudes. The extra pulses are analogous to the ones used for correcting qubit gates \cite{MagnusCorr17,MagnusCorr21} and can significantly decrease the protocol time allowing for states containing a large magnon number. 

For clarity in presentation, we consider the case of $g(n)\equiv g$ and 
\begin{equation}
	 \alpha(n)\equiv\frac{\alpha}{2}\left(n - 1\right),\label{def:alpha_num} 
\end{equation} 
while the calculations for the general case are presented in \ref{app:Transfer} and \ref{app:Gates}. The Hamiltonian, Eq. (\ref{eq:Ham_red}), becomes $\hat{H} = \Ham{s} + \Ham{d}(t)$, where the static part is 
\begin{equation}
	 \frac{\Ham{s}}{\hbar} = \omega_m\hat{m}^{\dagger}\hat{m} + \left(\omega_q - \frac{\alpha}{2}\left(\hat{q}^{\dagger}\hat{q} - 1\right)\right)\hat{q}^{\dagger}\hat{q} + g\left(\hat{m}\hat{q}^{\dagger} + \hat{m}^{\dagger}\hat{q}\right),\label{eq:HamiltonianSystem} 
\end{equation} 
and the dynamic part is 
\begin{equation}
	 \frac{\Ham{d}(t)}{\hbar} = \Delta(t)\hat{q}^{\dagger}\hat{q} + \tilde{\varepsilon}(t)\hat{q}^{\dagger} + \tilde{\varepsilon}^*(t)\hat{q} . 
\end{equation}

For a target magnon state $\ket{\psi}$, we want to find a set of operations $\hat{{\cal Q}}_j$ and $\hat{\mathcal{T}}_j$ satisfying (see Fig. \ref{Fig:ProtocolOper}) 
\begin{equation}
	 \ket{\Psi_0} = \ket{0,0},\ \ket{\tilde{\Psi}_j} = \hat{\mathcal{Q}}_j\ket{\Psi_j},\ \ket{\Psi_{j + 1}} = \hat{\mathcal{T}}_j\ket{\tilde{\Psi}_j},\ \ket{\Psi_{N_m}} = \ket{\psi,0}, 
\end{equation} 
where $\ket{\psi_1,\psi_2}$ refers to the magnon state $\ket{\psi_1}$ and transmon state $\ket{\psi_2}$. As before, starting from $\ket{\Psi_{N_m}}$, we inductively find $\{\hat{\mathcal{T}}_j,\hat{\mathcal{Q}}_j,\ket{\Psi_j}\}$ where the transfer operator ensures $\Op{j + 1,0}{\hat{\mathcal{T}}_j^{\dagger}}{\Psi_{j + 1}} = 0$ (removing a magnon) and the gate operation ensures 
\begin{equation}
	 \Op{j + 1 - s,s}{\hat{\mathcal{Q}}_j^{\dagger}}{\tilde{\Psi}_j} = 0, 
\end{equation} 
for $s\ge1$ (removing transmon excitations).  

\subsection{Transfer of excitation} 

In \ref{app:Transfer}, we derive $\hat{{\cal T}}_j$ satisfying the condition $\Op{j + 1,0}{\hat{\mathcal{T}}_j^{\dagger}}{\Psi_{j + 1}} = 0$, while we discuss the salient features of the derivation here. We start with the same ansatz as before [see Eq. (\ref{Def:Trans:TLS})] 
\begin{equation}
	 \hat{{\cal T}}^{{\rm ans}}\left(t_{{\rm on}},t_{{\rm off}}\right) = \exp\left[\frac{- i\Ham{s}t_{{\rm off}}}{\hbar}\right]\exp\left[ - \frac{i}{\hbar}\left(\Ham{s} + \Delta_0\hat{q}^{\dagger}\hat{q}\right)t_{{\rm on}}\right], 
\end{equation} 
except for the fact that the Hamiltonian includes all higher transmon levels. This ansatz can be simplified by noticing that the number of excitations $\hat{N} = \hat{m}^{\dagger}\hat{m} + \hat{q}^{\dagger}\hat{q}$ are conserved during the evolution, i.e. $\left[\hat{N},\hat{\cal T}^{{\rm ans}}\right] = \hat{0}$. An expression for $\hat{{\cal T}}^{{\rm ans}}$ restricted to the space of $\{\ket{j + 1 - s,s}\}_{s\ge0}$ can be found by numerically diagonalizing the two Hamiltonians, $\Ham{s}$ and $\Ham{s}+\Delta_0\hat{q}^{\dagger}\hat{q}$. Finally, we get a condition on $\{t_{{\rm on}},t_{{\rm off}}\}$ of the form 
\begin{equation}
	 \sum_{pqr}e^{i\lambda_{{\rm on}}^{(p)}t_{{\rm on}}}e^{i\lambda_{{\rm off}}^{(q)}t_{{\rm off}}}a_{pq}^{(r)}\inn{j + 1 - r,r}{\Psi_{j + 1}} = 0, 
\end{equation} 
where $\{\lambda_{{\rm on}}^{(p)},\lambda_{{\rm off}}^{(q)},a_{pq}^{(r)}\}$ can be found in terms of the eigenvectors and eigenvalues of the Hamiltonians [see \ref{app:Transfer}]. This equation can be numerically solved for $\{t_{{\rm on}},t_{{\rm off}}\}$. 

The above ansatz yields a solution in most cases, nevertheless, it can fail for some states, for example, when $\ket{\Psi_{j + 1}} = \ket{j + 1,0}$. Intuitively, when the coupling is on, the excitation from $\ket{j + 1,0}$ transfers to all $\ket{j + 1 - s,s}$ but with different oscillation frequencies $\sim\sqrt{g^2 + \frac{\alpha^2s^2(s - 1)^2}{4}}$. Unless these oscillation frequencies are commensurate, the amplitude of the wave-function in $\ket{j + 1,0}$ will stay non-zero. 

In cases where this ansatz fails, we can apply a second operation, 
\begin{equation}
	 \hat{{\cal T}}_j = \hat{{\cal T}}^{{\rm ans}}\left(t_{{\rm on},j}^{(2)},t_{{\rm off},j}^{(2)}\right)\hat{{\cal T}}^{{\rm ans}}\left(t_{{\rm on},j}^{(1)},t_{{\rm off},j}^{(1)}\right), 
\end{equation} 
where we choose $\{t_{{\rm on},j}^{(1)},t_{{\rm off},j}^{(1)}\}$ as the values found from a two-level approximation [see Eq. (\ref{eq:Gate_res:TLS})], and $\{t_{{\rm on},j}^{(2)},t_{{\rm off},j}^{(2)}\}$ are found by applying the above optimization to $\hat{{\cal T}}^{{\rm ans},\dagger}\left(t_{{\rm on},j}^{(1)},t_{{\rm off},j}^{(1)}\right)\ket{\Psi_{j + 1}}$. In principle, this process can be continued until a solution is found. However, in our simulations, we never had to go beyond a second step. 

\subsection{Transmon gates} 

As noted before, the application of $\hat{\mathcal{Q}}_j$ ensures that $\hat{\mathcal{Q}}_j^{\dagger}\ket{\tilde{\Psi}_j}$ has no excitations above the level $j$. The results are derived in \ref{app:Gates} and here, we discuss the salient features of the derivation. First, we want to apply a gate between transmon states $\ket{0}$ and $\ket{1}$, $\hat{\mathcal{Q}}_{01}^{{\rm ideal}}(\theta_j^{01},\phi_j^{01})$, to ``transfer'' the amplitude of $\ket{j,1}$ to $\ket{j,0}$, i.e. ensure $\Op{j,1}{\hat{{\cal Q}}_{01}^{{\rm ideal},\dagger}}{\tilde{\Psi}_j} = 0$, where $\{\theta_j^{01},\phi_j^{01}\}$ are given by 
\begin{equation}
	 \inn{j,1}{\tilde{\Psi}_j}\cos\theta_j^{01} + i\inn{j,0}{\tilde{\Psi}_j}\sin\theta_j^{01}e^{i\phi_j^{01}} = 0, 
\end{equation} 
which can be obtained using the same arguments as in the two-level approximation, see Eq. (\ref{eq:Gate_res:TLS}). To implement this with realistic signal, we use the ansatz 
\begin{equation}
	 \hat{\mathcal{Q}}_{01}(t) = \Texp{- \frac{i}{\hbar}\int_0^tdt\left(\Ham{s} - \hbar\varepsilon(\tau)e^{- i\omega_q\tau + i\phi_{{\rm app}}}\hat{q}^{\dagger} +{\rm h . c .}\right)}, 
\end{equation} 
where $\{\varepsilon(t),\phi_{{\rm app}}\}$ are to be determined. We can separate the phases by 
\begin{equation}
	 \hat{{\cal Q}}_{01}(t) \approx \exp\left[ - i\left(\omega_m\hat{m}^{\dagger}\hat{m} + \omega_q\hat{q}^{\dagger}\hat{q}\right)t\right]e^{i\phi_{{\rm app}}\hat{q}^{\dagger}\hat{q}}\hat{{\cal U}}_{01}(t)e^{- i\phi_{{\rm app}}\hat{q}^{\dagger}\hat{q}}, 
\end{equation} 
where we ignore the interaction between magnons and photons (they are far off-resonant) and 
\begin{equation}
	 \hat{{\cal U}}_{01}(t) = \Texp{i\int_0^td\tau\ \varepsilon(\tau)\left(\sum_{n = 0}^{\infty}\sqrt{n + 1}\ e^{- inA\tau}\ket{n + 1}\bra{n} + h . c . \right)} . 
\end{equation} 
 
In the two-level approximation, we can choose any $\varepsilon^{(0)}(t)$ satisfying $\int_0^{t_g}d\tau\varepsilon^{(0)}(\tau) = \theta_j^{01}$ and $\phi_{{\rm app}}^{(0)} = \phi_j^{01} - \omega_qt_g$. For concreteness, we choose 
\begin{equation}
	 \varepsilon^{(0)}(t) = \frac{\pi\theta_j^{01}}{2t_g}\sin\left(\frac{\pi t}{t_g}\right), 
\end{equation} 
where $t_g$ is a pre-determined gate time. In the two-level approximation, the time-dependence is given by 
\begin{equation}
	 \hat{{\cal U}}_{01}^{(0)}(t) = \Texp{i\int_0^td\tau\ \varepsilon^{(0)}(\tau)\left(\ket{1}\bra{0} + h . c . \right)} . 
\end{equation} 
As the term in the exponential commutes with itself at all times, we can remove the time-ordering operator and simplify $\hat{{\cal U}}_{01}^{(0)}$ to, 
\begin{equation}
	\fl \hat{{\cal U}}_{01}^{(0)}(t) = \cos\Theta(t)\left(\ket{0}\bra{0} + \ket{1}\bra{1}\right) + i\sin\Theta(t)\left(\ket{1}\bra{0} + \ket{0}\bra{1}\right) + \sum_{n = 2}^{\infty}\ket{n}\bra{n}, 
\end{equation} 
where $\Theta(t) = \int_0^t\varepsilon^{(0)}(\tau)d\tau$. 

However, a drive amplitude with a general form $\varepsilon(t)$ can induce transitions to higher transmon levels, e.g. transitions between $\ket{j,1}$ and $\ket{j,2}$. This is problematic as a significant transition amplitude $\Op{j,2}{\hat{{\cal Q}}_j^{\dagger}}{\tilde{\Psi}_j}$ implies excitation number $j + 2$, while we want $\hat{\mathcal{Q}}_j^{\dagger}\ket{\tilde{\Psi}_j}$ to have no excitations above $j$. Thus, we consider $\varepsilon(t) = \varepsilon^{(0)}(t) + \varepsilon^{(1)}(t)$, where $\varepsilon^{(1)}(t)$ are higher frequency components that cancel out higher order transitions, making $\hat{{\cal U}}_{01} \approx \hat{{\cal U}}_{01}^{(0)}$. 

The error in the transmon gate is given by 
\begin{equation}
	 \hat{{\cal U}}_{{\rm err}}(t) = \hat{{\cal U}}_{01}^{(0),\dagger}(t)\hat{{\cal U}}_{01}(t) = \Texp{i\int_0^td\tau\ \delta\hat{h}(\tau) - h . c .}, 
\end{equation} 
where 
\begin{equation}
	\fl \delta\hat{h}(\tau) = \hat{{\cal U}}_{01}^{(0),\dagger}(\tau)\left(\varepsilon(\tau)\sum_{n = 0}^{\infty}\sqrt{n + 1}\ e^{- inA\tau}\ket{n + 1}\bra{n} - \varepsilon^{(0)}(\tau)\ket{1}\bra{0}\right)\hat{{\cal U}}_{01}^{(0)}(\tau) . 
\end{equation} 
$\hat{{\cal U}}_{{\rm err}}$ cannot be found analytically. To linear order in the perturbation, we ignore the corrections due to non-commutativity, and impose 
\begin{equation}
	 \int_0^{t_g}d\tau\ \delta\hat{h}(\tau) = \hat{0} . 
\end{equation} 
This in turn implies the following set of equations that the drive amplitude $\varepsilon(t)$ must satisfy 
\begin{eqnarray}
	 \int_0^{t_g}\left(\varepsilon(\tau) - \varepsilon^{(0)}(\tau)\right)d\tau &= 0,\label{eq:Conds1}\\
	 \int_0^{t_g}\varepsilon(\tau)e^{iA\tau\pm i\Theta(t)}d\tau &= 0,\label{eq:Conds2}\\ n\ge2:\ \int_0^{t_g}\varepsilon(\tau)e^{in\tau}d\tau &= 0 . \label{eq:Conds3} 
\end{eqnarray} 
The first condition is a no-bias condition ensuring that $\int_0^{t_g}\varepsilon(\tau)d\tau = \theta_j^{01}$. The second condition, with $\Theta(t) = \int_0^td\tau\varepsilon^{(0)}(\tau)$ ensures that the transitions $\ket{1}\leftrightarrow\ket{2}$ cancel out at the end of the gate. Similarly, the transitions $\ket{n}\leftrightarrow\ket{n + 1}$ are cancelled by the third condition. In our simulations, we find that for typical parameters, the population beyond level $2$ is negligible, so we can ignore the conditions for $n\ge3$. 

The above conditions leave sufficient freedom in choosing $\varepsilon(t)$. In what follows, we choose $t_g = 3\pi/\alpha$ and 
\begin{equation}
	 \varepsilon(t) = \sum_{\mu = 0}^4\varepsilon_{\mu}\sin\left(\frac{\left(2\mu + 1\right)\pi t}{t_g}\right), 
\end{equation} 
where the parameters $\varepsilon_{\mu}$ are governed by an underdetermined set of linear equations that can be found from the above conditions [shown explicitly in \ref{app:Gates}]. Given $\varepsilon(t)$, we can find $\phi_{{\rm app}}$ by minimizing $\left|\Op{j,1}{\hat{{\cal Q}}_{01}^{\dagger}}{\tilde{\Psi}_j}\right|$ as discussed in \ref{subapp:Phase_Gate}. 

We point out that although we choose a linear approximation, the corrections are not always small. However, in our simulations we find that the error matrix $\hat{{\cal U}}_{{\rm err}}$ is dominated by extraneous phases for all transmon levels. One might think that these can be removed by adding higher order corrections, however, we observed that such higher order corrections typically shifts the errors from $\ket{2}$ to $\ket{n\ge3}$ instead of removing them \footnote{A similar correction scheme was followed in \cite{MagnusCorr17,MagnusCorr21} by ignoring levels higher than $\ket{2}$ where this effect was not observed.}. Thus, we account for these extra phases by modifying the next steps of the induction. 

If $\left|\Op{j - 1,2}{\hat{{\cal Q}}_{01}^{\dagger}}{\tilde{\Psi}_j}\right|$ is significant (we take the threshold as $0 . 1$ in our simulations), a second gate between $\ket{1}$ and $\ket{2}$, say $\hat{{\cal Q}}_{12}^{{\rm ideal}}(\theta_j^{12},\phi_j^{12})$, can be applied to ensure that $\Op{j - 1,2}{\hat{{\cal Q}}_{12}^{{\rm ideal},\dagger}\hat{{\cal Q}}_{01}^{\dagger}}{\tilde{\Psi}_j} = 0$, with 
\begin{equation}
	 \Op{j,1}{\hat{{\cal Q}}_{01}^{\dagger}}{\tilde{\Psi}_j}\cos\theta_j^{12} + i\Op{j,0}{\hat{{\cal Q}}_{01}^{\dagger}}{\tilde{\Psi}_j}\sin\theta_j^{12}e^{i\phi_j^{12}} = 0 . 
\end{equation} 
Essentially, the same arguments presented above hold for the gate $\hat{{\cal Q}}_{12}$ defined by 
\begin{equation}
	\fl \hat{\mathcal{Q}}_{12}\left[\varepsilon(t),\phi_{{\rm app}}\right] = \Texp{- \frac{i}{\hbar}\int_0^{t_g}dt\left(\Ham{s} - \frac{\hbar\varepsilon(t)}{\sqrt{2}}e^{- i\left(\omega_q - A\right)t + i\phi_{{\rm app}}}\hat{q}^{\dagger} +{\rm h . c .}\right)}, 
\end{equation} 
where we can choose 
\begin{equation}
	 \varepsilon^{(0)}(t) = \frac{\pi\theta_j^{12}}{2t_g}\sin\left(\frac{\pi t}{t_g}\right) 
\end{equation} 
and $\varepsilon(t)$ satisfying the same conditions as before. 

The levels $\ket{n\ge3}$ are found to be approximately unoccupied, so we do not need to apply more gates. 

\section{Results\label{sec:Results}} 

To evaluate the above algorithm, we numerically solve the time evolution that implements the protocol for the density matrix, using QuTiP \cite{QuTiP1,QuTiP2}, satisfying 
\begin{equation}
	\fl \frac{d\hat{\rho}}{dt} =  - \frac{i}{\hbar}\left[\Ham{s} + \Ham{d}(t),\hat{\rho}(t)]\right] + \sum_n\frac{1}{2}\left[2\hat{C}_n\hat{\rho}(t)\hat{C}_n^{\dagger} - \hat{\rho}(t)\hat{C}_n^{\dagger}\hat{C}_n - \hat{C}_n^{\dagger}\hat{C}_n\hat{\rho}(t)\right] . 
\end{equation} 
For dissipation, we take the Lindblad operators pertaining to magnon decay $\hat{C}_1 = \sqrt{\kappa_m}\hat{m}$, transmon decay $\hat{C}_2 = \sqrt{\kappa_q}\hat{q}$ and transmon dephasing $\hat{C}_3 = \sqrt{\gamma_q}\hat{q}^{\dagger}\hat{q}$. We did not include magnon dephasing as it has not yet been experimentally observed, although there are theoretical reasons for pure dephasing of magnons \cite{Dephasing}. We choose typical parameters $\omega_m = 2\pi\times\SI{6}{\giga\hertz}$, $\omega_q = 2\pi\times\SI{5}{\giga\hertz}$, $\kappa_m = 10^{- 4}\omega_m$, $\kappa_q = \gamma_q = 10^{- 4}\omega_q$, $A = 2\pi\times\SI{300}{\mega\hertz}$, and coupling $g = 2\pi\times\SI{25}{\mega\hertz}$. For a given target state $\ket{\psi_T}$, the algorithm described in the previous sections gives the signals $\Delta[\psi_T](t)$ and $\tilde{\varepsilon}[\psi_T](t)$, along with the protocol time $T_{{\rm tot}}[\psi_T]$. Starting from the ground state, $\ket{0,0}$, we numerically solve the above Lindblad equation with the given signals, $\Delta$ and $\tilde{\varepsilon}$, and find the reduced density matrix for the magnon, $\hat{\rho}_{{\rm mag}}[\psi_T] ={\rm Tr}_{\hat{q}}\left[\hat{\rho}(t[\psi_T])\right]$: 
\begin{equation}
	 \ket{\psi_T} \oset{\rm protocol}{\longrightarrow} \{\Delta(t),\tilde{\varepsilon}(t)\} \oset{\rm simulation}{\longrightarrow}\hat{\rho}_{{\rm mag}} . 
\end{equation} 
Finally, we calculate the fidelity of state generation $F[\psi_T] = \Op{\psi_T}{\hat{\rho}_{{\rm mag}}[\psi_T]}{\psi_T}$ which can be interpreted as the probability of finding the magnetization in the target state $\ket{\psi_T}$. 

We compare the Wigner function of the target state $\ket{\psi_T}$ and the achieved state $\hat{\rho}_{{\rm mag}}[\psi_T]$ for some special cases. The Wigner function is defined as 
\begin{equation}
	 W(\beta,\hat{\rho}) = \int\frac{d^2\gamma}{\pi}{\rm Tr}\left[\hat{\rho}\exp\left[\left(\hat{m} - \beta\right)^{\dagger}\gamma - \gamma^*\left(\hat{m} - \beta\right)\right]\right] . 
\end{equation} 
The Wigner function can be interpreted as follows. Define a symmetric product $\left\{\hat{m}^{\dagger,p}\hat{m}^q\right\}_S$ as an average over all permutations of $p$ $\hat{m}^{\dagger}$'s and $q$ $\hat{m}$'s. For example, 
\begin{equation}
	 \left\{\hat{m}^{\dagger,2}\hat{m}\right\}_S = \frac{\hat{m}^{\dagger,2}\hat{m} + \hat{m}^{\dagger}\hat{m}\hat{m}^{\dagger} + \hat{m}\hat{m}^{\dagger,2}}{3} . 
\end{equation} 
In a given state $\hat{\rho}$, the average of the above operator is given by \cite{Cahill1,Cahill2} 
\begin{equation}
	{\rm Tr}\left[\hat{\rho}\left\{\hat{m}^{\dagger,p}\hat{m}^q\right\}_S\right] = \int\frac{d^2\beta}{\pi}\beta^{*,p}\beta^qW(\beta,\hat{\rho}) . 
\end{equation} 
We can convert this into expectation values of the magnetization operators $\hat{M}_x$ and $\hat{M}_y$ via the substitution $2\ZPF\hat{m}\rightarrow\hat{M}_x - i\hat{M}_y$ (note that we are using the dressed operator $\hat{m}$ instead of the bare magnetization operator $\hat{M}$ as the distinction between them is small) getting 
\begin{equation}
	{\rm Tr}\left[\hat{\rho}\left\{\hat{M}_x^p\hat{M}_y^q\right\}_S\right] = \int\frac{dM_xdM_y}{\pi}M_x^pM_y^q\ \tilde{W}\left(M_x,M_y,\hat{\rho}\right), 
\end{equation} 
where $\tilde{W}\left(M_x,M_y . \hat{\rho}\right) = W\left(\frac{M_x - iM_y}{2\ZPF},\hat{\rho}\right)$. The above equation suggests an interpretation of $\tilde{W}$ as a quasi-classical probability distribution. However, note that $\tilde{W}$ is not always positive and a negative $\tilde{W}$ implies a lack of a probabilistic interpretation considered to be a signature of non-classicality. The unit $\ZPF$, defined in Eq. (\ref{Def:ZPF}), is given by 
\begin{equation}
	 \frac{\ZPF}{M_s} = 8\times10^{- 6}\sqrt{\frac{\SI{1}{\micro\meter\cubed}}{V_m}}, 
\end{equation} 
where $V_m$ is the volume of the magnet. This is a very small angle but it can be measured via the magnet's coupling to the transmon \cite{MagDet2,MagDet3}. 

\begin{figure*}
	\includegraphics[width=1\columnwidth]{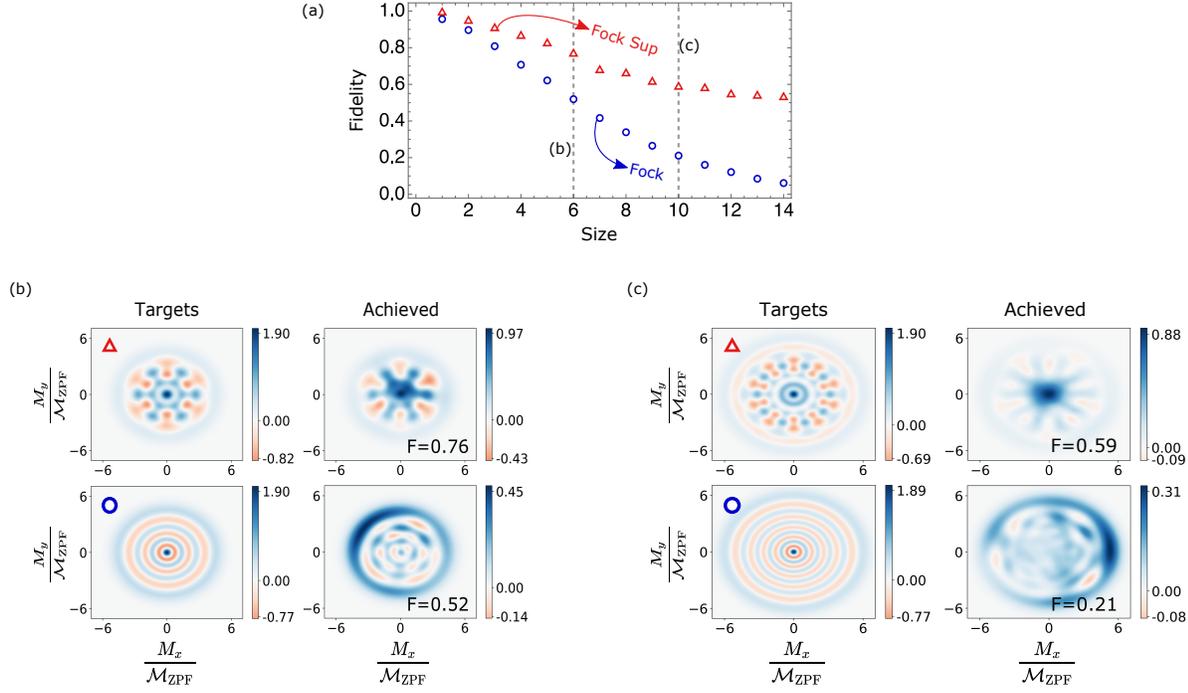}
	\caption{(a) Fidelity of generation for Fock states $\ket{n}$ (blue circles) and superpositions between the vacuum and Fock states $(\ket{0}+\ket{n})/\sqrt{2}$ (red triangles) as a function of $n$. Wigner functions of target and achieved state are plotted for $n=6$ in (b) and $n=10$ in (c) with fidelity (F) written.}\label{Fig:Focks}
\end{figure*}

In Fig. \ref{Fig:Focks}, we consider the case of Fock states $\ket{\psi_{{\rm Fock}}(n)} = \ket{n}$ and their superposition with the ground state $\ket{\psi_{{\rm Sup}}(n)}\propto\ket{0} + \ket{n}$ where $n\in\{1,2,\dots,14\}$. The fidelity decreases with increasing $n$. This is caused by two factors. The dominant factor is an increase in the protocol time with the number of magnons leading to higher decay. Secondly, each gate introduces a small amount of error, including residual transfer to higher levels, that compounds with the number of gates, linearly increasing with $n$. It appears that Fock states are more difficult to generate than the superposition states, which could be due to a larger average number of magnons in $\ket{\psi_{{\rm Fock}}}$, compared to $\ket{\psi_{{\rm Sup}}}$, leading to a faster decay. For $n\in\{6,10\}$ and $T\in\{{\rm Fock},{\rm Sup}\}$, we also show Wigner functions of the target state, $\tilde{W}\left(M_x,M_y,\ket{\psi_T(n)}\bra{\psi_T(n)}\right)$ and the achieved state $\tilde{W}\left(M_x,M_y,\hat{\rho}_{{\rm mag}}\left[\psi_T(n)\right]\right)$. The essential feature of $\ket{\psi_{{\rm Sup}}(n)}$ are the azimuthal fringes of order $n$, which is well captured for $n = 6$ and weakly reproduced for $n = 10$. The essential feature of $\ket{\psi_{{\rm Fock}}(n)}$ are the $n$ circular fringes, that are destroyed by dephasing. 

\begin{figure*}
	\includegraphics[width=1\columnwidth]{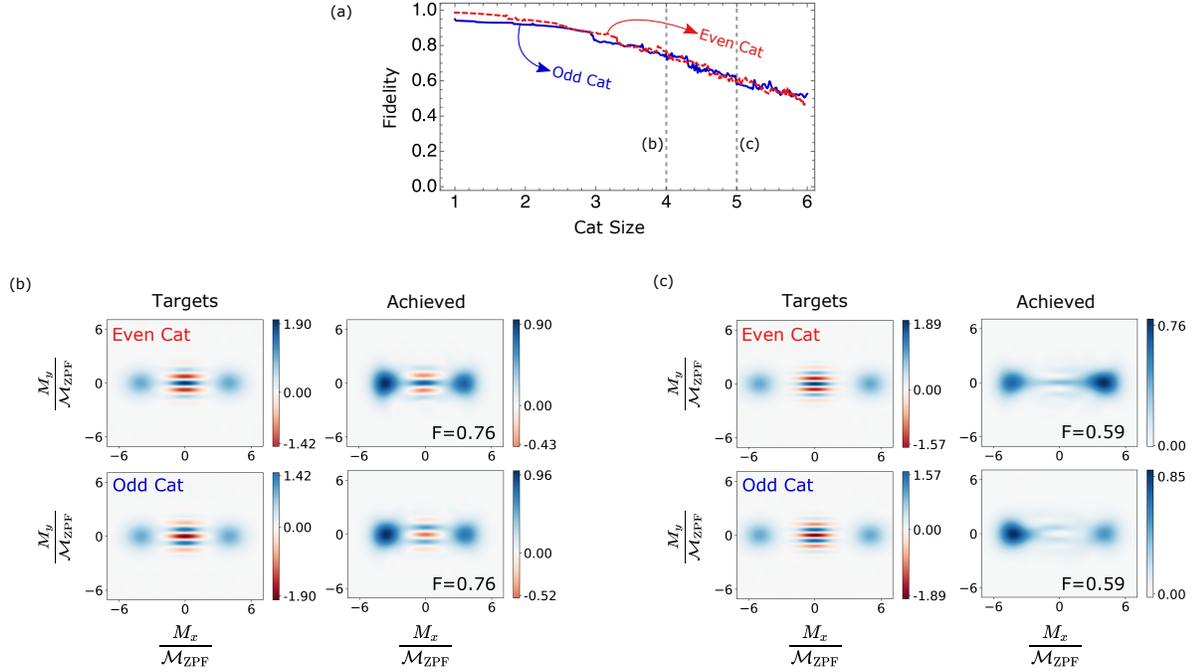}
	\caption{(a) Fidelity of generation for cat states $\ket{\xi/2} \pm \ket{-\xi/2}$ as a function of cat size. Wigner functions of target and achieved state are plotted for $\xi=4$ in (b) and $\xi=5$ in (c) with fidelity (F) written.}\label{Fig:Cats}
\end{figure*}

A superposition of two semi-classical states, known as a cat state, is useful for error correction protocols because of their insensitivity to particle loss noise \cite{Ofek_QEC}, particularly as carriers of information \cite{Jeong_QCCat,Ralph_QCCat,Mirrahimi_QCCat} and in metrology \cite{MunroMetro,Knott_PracticalQMetro,Huang_QuantumMetro,RalphMetro}. In case of a magnet, we can define semi-classical states as the minimum-uncertainty states with isotropic quantum fluctuations, given by coherent states \cite{Glauber} 
\begin{equation}
	 \ket{\xi} = e^{- \left|\xi\right|^2/2}\sum_{n = 0}^{\infty}\frac{\xi^n}{\sqrt{n!}}\ket{n}, 
\end{equation} 
for an arbitrary complex number $\xi$. A coherent state has average $\avg{\hat{m}} = \xi$ and can be interpreted as the quantum state `closest' to the classical state with magnetization $M_x = 2\ZPF\mathrm{Re}\left(\xi\right)$ and $M_y =  - 2\ZPF\mathrm{Im}\left(\xi\right)$. An `odd' cat state given by $\ket{\psi_{{\rm odd}}(\xi)}\propto\ket{\xi/2} - \ket{- \xi/2}$ has two semi-classical components separated by the magnetization $2\ZPF\left|\xi\right|$; the same holds for an `even' cat state $\ket{\psi_{{\rm even}}(\xi)\propto\ket{\xi/2} + \ket{- \xi/2}}$. Fig. \ref{Fig:Cats} shows the fidelity of generation as a function of $\xi$. The Wigner function for an ideal cat state has two blobs at the expected peaks along with fringes in the center. The presence of the fringes signify the quantum coherence between the two semi-classical components. In contrast, the Wigner function for a classical distribution, $\hat{\rho}_{{\rm cl}}(\xi)\propto\ket{\xi/2}\bra{\xi/2} + \ket{- \xi/2}\bra{- \xi/2}$ would still have the blobs but no fringes. 

In the achieved states, we find such fringes for cat sizes up to 4 until which the fidelities are not too low. At cat size 5, we see that the achieved state are closer to a classical probability distribution [they have $>0 . 85$ fidelity with the corresponding classical states $\hat{\rho}_{{\rm cl}}(\xi=5)]$. Differences in the odd and cat states for low sizes can be explained on the basis that odd cats have larger number of magnons, $\frac{\xi^2}{4}\frac{1 + e^{- \xi^2/2}}{1 - e^{- \xi^2/2}}$, than even cats, $\frac{\xi^2}{4}\frac{1 - e^{- \xi^2/2}}{1 + e^{- \xi^2/2}}$, an effect that is negligible for $\xi>2$. Fidelity for generation of any other state can be found using the available codes \cite{Codes}.

\section{Improvements\label{sec:Improvements}} 

Here, we focussed on adapting and characterizing the protocol for a system with a magnet and a transmon. There are several methods by which the protocol can be improved, as we now elaborate. 

\emph{Pre-processing:} We considered only a specific type of operations, where the transmon is excited and the excitation is transferred to the magnons. With the help of the MW cavity, one can also apply other operations, in particular, a displacement given by $\hat{D}(\alpha) = e^{\alpha\hat{m}^{\dagger} - \alpha^*\hat{m}}$ caused by a coherent excitation of the magnet by MWs. This can be used to decrease the time of the protocol as one can choose an appropriate $\alpha$, such that $\hat{D}^{\dagger}(\alpha)\ket{\psi_T}$ has fewer magnons than $\ket{\psi_T}$. For example, if $\Op{\psi_T}{\hat{m}}{\psi_T} = C$, then 
\begin{equation}
	 \Op{\psi_T}{\hat{D}(C)\hat{m}^{\dagger}\hat{m}\hat{D}^{\dagger}(C)}{\psi_T} = \Op{\psi_T}{\hat{m}^{\dagger}\hat{m}}{\psi_T} - \left|C\right|^2, 
\end{equation} 
implying that $\hat{D}^{\dagger}(C)\ket{\psi_T}$ has fewer average magnons than $\ket{\psi_T}$. A better $\alpha$ can be found numerically that decreases the highest amplitude in $\ket{\psi_T}$ to within some given tolerance. Then, we run the protocol to create $\hat{D}^{\dagger}(\alpha)\ket{\psi_T}$ and apply a MW pulse to displace keeping a high fidelity. If there are other operations that can be easily performed, they can be used for pre-processing to decrease the size. 

\emph{High fidelity gates: }In our analysis, we find that non-ideal transmon gates are responsible for a significant part of fidelity loss which increases linearly with the number of gates necessary to generate the target state. We took a simple case of correcting at first order perturbation similar to \cite{MagnusCorr17,MagnusCorr21}. Simply adding higher order terms in a Dyson or a Magnus expansion does not necessarily improve gate fidelities. When we correct for second order term in Magnus expansion, we find that we can decrease the transitions $\ket{0}\leftrightarrow\ket{2}$ and $\ket{1}\leftrightarrow\ket{2}$ significantly, but at the expense of a high $\ket{0}\leftrightarrow\ket{3}$ transition probability \footnote{In Refs. \cite{MagnusCorr17,MagnusCorr21}, only a three level system was considered which does not have this problem}. This gets corrected only at the third order in Magnus expansion, at which point the $\ket{0}\leftrightarrow\ket{4}$ transition might play a role. However, there is a rich literature on speeding up gates to $\sim2\pi/\alpha$ with low loss of fidelity, such as DRAG \cite{DRAG,Leakage_Exp2} or optimal control \cite{ashhabSpeed,Leakage_Exp,Gate_OCT,Gate_OCT2}, that can not only speed up but also improve the accuracy of gates. 

\emph{Combining steps: }We followed a semi-analytical approach where we systematically add an excitation one by one. In principle, some steps can be combined, e.g. exciting the transmon and transferring the excitation to magnon can occur simultaneously. This becomes infeasible to handle analytically but relevant pulses can, in principle, be found using optimal control theory. A similar case has been discussed in the two-level approximation \cite{RojanOCT} for a general harmonic oscillator states with $2 - 3$ quanta. Although this approach should result in the least possible errors, it is computationally very expensive and scaling such an approach to high magnon numbers \cite{OCT_Toolkit,Leakage_Exp} would still require a good initial guess for time-dependent functions describing the transmon driving and the transmon frequency.  

\section{Conclusions\label{sec:Conclusions}} 

We proposed and numerically demonstrated a protocol to generate an arbitrary quantum state of the magnetization using its coupling to a superconducting transmon, which is a weakly anharmonic oscillator. The protocol involves repeatedly applying transmon gates and transferring an excitation to the magnons. As longer protocol times lead to higher dissipation, we can improve fidelity of generation by reducing the gate time, $t_g$, and the transfer time, $t_{{\rm on}}$, in the notation of Secs. \ref{sec:Gen:TLS} and \ref{sec:Gen:Transmon}. Fast gate pulses with duration $\sim2\pi/\alpha$ ($\alpha$ is defined in Eq. (\ref{def:alpha_num})) would imply leakage of excitation into higher transmon levels that can be reduced by modifying the input pulses \cite{Rol19,ashhabSpeed,Leakage_Exp,DRAG}. We use correction pulses by removing errors at first order in a Magnus expansion, similar to Refs. \cite{MagnusCorr17,MagnusCorr21}, bringing the gate time down to $3\pi/\alpha$. Correcting for second order term in the expansion does not seem to decrease errors, but rather transfer the population to higher levels. The transfer time, $t_{{\rm on}}$, can be reduced with a higher magnon-transmon coupling. As the magnon-MW coupling can reach very high values, it is likely to be experimentally feasible to have high magnon-transmon couplings. However, a coupling $g\sim\alpha$ causes leakage of magnons into higher transmon levels leading to loss of fidelities, so we choose $g<0 . 1\alpha$. For such couplings, the leakage is small and can be incorporated by renormalizing the duration of the coupling, $t_{{\rm on},j}$. 

The main source of errors is the magnon dissipation putting a limit on the size of the states. We find that the protocol faithfully produces states up to a largest occupation of $\sim9$ magnons and an average of $\sim4$ magnons. We note that our protocol creates states in dressed magnons, given by $\hat{m}$ instead of bare magnons $\hat{M}$. As discussed in \ref{app:Ham_eff}, for a sufficiently large detuning between MWs and magnons, the distinction is small. We demonstrated our protocol for cases of Fock states $\ket{n}$, Fock superpositions $\ket{0} + \ket{n}$, and cat states $\ket{\alpha}\pm\ket{- \alpha}$. While our protocol gives good fidelities for states of moderate sizes, we expect that advanced numerical techniques such as optimal quantum control \cite{OCT_Toolkit,Leakage_Exp,Gate_OCT} can improve the state generation (likely by reducing the protocol time) by using our protocol as an initial guiding guess. 

With a setup to generate an arbitrary magnetization, there still lies a question of how to experimentally measure it, which we leave to future work. The figures can be generated via the publicly available codes \cite{Codes}.

\section*{Acknowledgements} 

The authors thank H. Ribeiro for useful discussions. We acknowledge funding from the Max Planck Society and from the Deutsche Forschungsgemeinschaft (DFG, German Research Foundation) through Project-ID 429529648--TRR 306 QuCoLiMa (\textquotedblleft Quantum Cooperativity of Light and Matter\textquotedblright ). 

\appendix 

\section{Derivation of effective Hamiltonian\label{app:Ham_eff}} 

In this appendix, we derive the effective Hamiltonian for a magnet coupled to a transmon indirectly via a MW cavity.  

\subsection{Microwaves} 

The MW cavity is modeled by the classical Hamiltonian density 
\begin{equation}
	 \ClHam{mw} = \frac{\epsilon_0\left|\bi{E}\right|^2 + \mu_0\left|\bi{H}\right|^2}{2} . 
\end{equation} 
It contains several modes labelled by $r$ giving the quantization 
\begin{equation}
	 \bi{H}(\bi{r})\rightarrow\sum_r\bi{H}_r(\bi{r})\hat{A}_r + \bi{H}_r^*(\bi{r})\hat{A}_r^{\dagger}, 
\end{equation} 
and similar holds for $\bi{E}$. Here, $\bi{H}_r$ and $\hat{A}_r$ are respectively the mode profile and the annihilation operator of mode $r$. With appropriate normalization, this gives the Hamiltonian, 
\begin{equation}
	 \Ham{mw} = \hbar\sum_r\Omega_{a,r}\hat{A}_r^{\dagger}\hat{A}_r, 
\end{equation} 
where $\Omega_{a,r}$ is the resonance frequency of mode $r$. 

\subsection{Magnet} 

If the magnet is much smaller than MW wavelengths ($\sim\si{\centi\meter}$), then only the total magnetization $\bi{M}$ couples to the MWs. Under a DC field along $\bi{z}$ with magnitude $H_{{\rm app}}$, its classical Hamiltonian density is 
\begin{equation}
	 \ClHam{m,mw} =  - \mu_0H_{{\rm app}}M_z - \mu_0\bi{M}\cdot\bi{H}, 
\end{equation} 
where $\bi{H}$ is the magnetic field due to the cavity. The magnetization is quantized by the Holstein-Primakoff transformation \cite{HolPrim,StanPrabh}, given by a form 
\begin{equation}
	 M_x - iM_y\rightarrow2\ZPF\hat{M}\sqrt{\frac{1 - \beta\hat{M}^{\dagger}\hat{M}}{1 - \beta}}, 
\end{equation} 
where $\ZPF$ and $\beta$ are to be found. We know that a spin in an applied field undergoes a Larmor precession with frequency $\Omega_m = \gamma\mu_0H_{{\rm app}}$ where $\gamma$ is the absolute value of the gyromagnetic ratio. This implies that the Kittel mode should have the resonance frequency, i.e. after quantization 
\begin{equation}
	 - \mu_0V_mH_{{\rm app}}M_z\rightarrow\hbar\Omega_m\hat{M}^{\dagger}\hat{M} . 
\end{equation} 
This implies $M_z = \tilde{M}_s - M_{{\rm sing}}M^{\dagger}M$ where $\tilde{M}_s$ (to be found) is the saturation magnetization reduced by zero point fluctuations and $M_{{\rm sing}} = \gamma\hbar/V_m$ is the magnetization reduction because of one (delocalized) spin flip. Using $M_x^2 + M_y^2 + M_z^2 = M_s^2$, we find the relations 
\begin{equation}
	 \tilde{M}_s^2 + 2\ZPF^2 = M_s^2,\ \ZPF = \sqrt{\frac{\tilde{M}_sM_{{\rm sing}}}{2}},\ M_{{\rm sing}}^2 = \frac{4\ZPF^2\beta}{1 - \beta} . 
\end{equation} 
This is solved explicitly by 
\begin{equation}
	 \tilde{M}_s = \sqrt{M_s^2 + \frac{M_{{\rm sing}}^2}{2}} - \frac{M_{{\rm sing}}}{2},\ \beta = \frac{M_{{\rm sing}}}{2\tilde{M}_s + M_{{\rm sing}}} 
\end{equation} 
Typically $M_{{\rm sing}}/M_s<10^{- 10}$, so we can resort to the leading order corrections, 
\begin{equation}
	 \tilde{M}_s \approx M_s - \frac{M_{{\rm sing}}}{2},\ \ZPF \approx \sqrt{\frac{M_sM_{{\rm sing}}}{2}},\ \beta \approx \frac{M_{{\rm sing}}}{M_s} . 
\end{equation} 
For $\avg{\hat{M}^{\dagger}\hat{M}}\ll1/\beta\sim10^{10}$, we can simply substitute 
\begin{equation}
	 M_x - iM_y\rightarrow2\ZPF\hat{M},\ M_z\rightarrow M_s - \frac{2\ZPF^2}{M_s}\hat{M}^{\dagger}\hat{M} . 
\end{equation} 
This gives the magnon Hamiltonian (coupled to the cavity) 
\begin{equation}
	 \Ham{m,mw} = \hbar\Omega_m\hat{M}^{\dagger}\hat{M} + \sum_rG_{m,r}\hat{M}^{\dagger}\hat{A}_r + G_{m,r}^*\hat{M}\hat{A}_r^{\dagger}, 
\end{equation} 
where $\Omega_m = \gamma\mu_0H_{{\rm app}}$ and 
\begin{equation}
	 \hbar G_{m,r} =  - \mu_0\ZPF\left(H_{x,r} - iH_{y,r}\right) . 
\end{equation} 
We assumed rotating wave approximation which is valid when $G\ll\Omega_m + \Omega_a$. 

\subsection{Transmon} 
We consider a flux-tunable transmon that consists of two Josephson junctions in parallel forming a SQUID loop. Its Hamiltonian is \cite{Koch_Transmon,Blais_RMP} 
\begin{equation}
	 \Ham{tr,mw} = E_C\left(\hat{n} + \hat{n}_{{\rm mw}}\right)^2 - E_{J1}\cos\hat{\phi}_1 - E_{J2}\cos\hat{\phi}_2 . 
\end{equation} 
Here, $\hat{n}$ is the number of quanta exchanged between the two junctions, $\hat{n}_{{\rm mw}}$ is the number of quanta excited by MWs, and $\hat{\phi}_{1,2}$ are the fluxes through each junction. Under weak coupling, we have $\hat{n}_{{\rm mw}}\propto\left|\bi{E}\right|$. Flux quantization constraints $\hat{\phi}_1 - \hat{\phi}_2 = 2\pi m + 2\pi\Phi/\Phi_0$ where $m$ is an arbitrary integer, $\Phi$ is the flux through the SQUID loop, and $\Phi_0 = h/2e$ is the superconducting flux quantum. Defining $2\hat{\phi} = \hat{\phi}_1 + \hat{\phi_2}$, 
\begin{equation}
	 \Ham{tr} = E_C\hat{n}^2 - E_{J,{\rm eff}}\cos\left(\hat{\phi} + \phi_0\right) . 
\end{equation} 
where the phase is 
\begin{equation}
	 \phi_0 = \tan^{- 1}\left(\frac{E_{J1} - E_{J2}}{E_{J1} + E_{J2}}\tan\frac{\pi\Phi}{\Phi_0}\right) + n\pi 
\end{equation} 
and the effective junction energy is 
\begin{equation}
	 E_{J,{\rm eff}}(\Phi) = \sqrt{\left(E_{J1} + E_{J2}\right)^2\cos^2\frac{\pi\Phi}{\Phi_0} + \left(E_{J1} - E_{J2}\right)^2\sin^2\frac{\pi\Phi}{\Phi_0}} . 
\end{equation} 
In the transmon limit $E_{J,{\rm eff}}\gg E_C$, the fluctuations in $\hat{\phi} + \phi_0$ are small, so we can expand the cosine and get an approximately quadratic Hamiltonian. This suggests the quantization, 
\begin{equation}
	 \hat{n} = \left(\frac{E_{J,{\rm eff}}}{2E_C}\right)^{1/4}\frac{\hat{Q} - \hat{Q}^{\dagger}}{\sqrt{2}i},\ \hat{\phi} + \phi_0 = \left(\frac{2E_C}{E_{J,{\rm eff}}}\right)^{1/4}\frac{\hat{Q} + \hat{Q}^{\dagger}}{\sqrt{2}}, 
\end{equation} 
that reduces the Hamiltonian to 
\begin{equation}
	 \Ham{q,mw} = \hbar\Omega_q\hat{Q}^{\dagger}\hat{Q} - \hbar\frac{A}{2}\hat{Q}^{\dagger}\hat{Q}\left(\hat{Q}^{\dagger}\hat{Q} - 1\right) + \hbar\sum_r\left(G_{q,r}\hat{Q}^{\dagger}\hat{A}_r + G_{q,r}^*\hat{Q}\hat{A}_r^{\dagger}\right), 
\end{equation} 
ignoring higher order terms. Here $\hbar\Omega_q = \sqrt{2E_CE_{J,{\rm eff}}} - E_C/4$ and $\hbar A = E_C/4$. 

\subsection{Dressed magnons and transmons}

The dressed states are found by diagonalizing the total Hamiltonian $\hat{H} = \Ham{0} + \Ham{int}$ where 
\begin{equation}
	 \frac{\Ham{0}}{\hbar} = \sum_r\Omega_{a,r}\hat{A}_r^{\dagger}\hat{A}_r + \Omega_m\hat{M}^{\dagger}\hat{M} + \Omega_q\hat{Q}^{\dagger}\hat{Q} - \frac{A}{2}\hat{Q}^{\dagger}\hat{Q}\left(\hat{Q}^{\dagger}\hat{Q} - 1\right), 
\end{equation} 
and the interaction 
\begin{equation}
	 \frac{\Ham{int}}{\hbar} = \sum_r\left(G_{m,r}\hat{M}^{\dagger}\hat{A}_r + G_{m,r}^*\hat{M}\hat{A}_r^{\dagger} + G_{q,r}\hat{Q}^{\dagger}\hat{A}_r + G_{q,r}^*\hat{Q}\hat{A}_r^{\dagger}\right) . 
\end{equation} 
However, exact diagonalization is not possible analytically, so we resort to a perturbative treatment, in particular a Schrieffer-Wolff (SW) transformation \cite{SW_Review}. The SW transformation perturbatively gives the rotation operator $\hat{S} =  - \hat{S}^{\dagger}$ such that $e^{- \hat{S}}\hat{H}e^{\hat{S}}$ is diagonal to a given order. We find $\hat{S}$ to linear order in coupling constants $G$ that gives the rotated Hamiltonian, 
\begin{equation}
	 e^{- \hat{S}}\hat{H}e^{\hat{S}} = \Ham{0} + \Ham{int} + \left[\Ham{0},\hat{S}\right] + \left[\Ham{int},\hat{S}\right] + \frac{1}{2}\left[\left[\Ham{0},\hat{S}\right],\hat{S}\right] + O\left(G^3\right) . 
\end{equation} 
To remove first order corrections, we choose 
\begin{equation}
	 \hat{S} = \sum_r\left(\frac{G_{m,r}}{\Omega_{a,r} - \Omega_m}\hat{A}_r\hat{M}^{\dagger} + G_{q,r}\chi_r\left[\hat{Q}^{\dagger}\hat{Q}\right]\hat{A}_r\hat{Q}^{\dagger}\right) - h . c . 
\end{equation} 
where the transmon susceptibility is 
\begin{equation}
	 \chi_r\left[x\right] = \frac{1}{\Omega_{a,r} - \Omega_q - \alpha\left(x - 1\right)}, 
\end{equation} 
ensuring $\Ham{int} + \left[\Ham{0},\hat{S}\right] = \hat{0}$. Note that we need $\left\Vert \hat{S}\right\Vert \ll1$ that implies a limit on the number of excitations $\{N_{a,r},N_m,N_q\}$ s.t. for all $r$, 
\begin{equation}
	 G_{m,r}\sqrt{N_{a,r}N_m}\ll\Omega_{a,r} - \Omega_m,\ G_{q,r}\sqrt{N_{a,r}N_q}\ll\Omega_{a,r} - \Omega_q 
\end{equation} 
are satisfied. This gives the rotated Hamiltonian, 
\begin{equation}
	 e^{- \hat{S}}\hat{H}e^{\hat{S}} = \Ham{0} + \frac{1}{2}\left[\Ham{int},\hat{S}\right] + O\left(G^3\right) . 
\end{equation} 
As there are four terms in both $\Ham{int}$ and $\hat{S}$, a total of $16$ commutators are required. If we ignore all non-resonant terms, such as $\hat{A}_r^2\hat{Q}^{\dagger,2}$, and use the Hermiticity of $\Ham{int}$ we are left with 
\begin{equation}
	\fl \frac{1}{2\hbar}\left[\Ham{int},\hat{S}\right] = \sum_{rs}\left[\left(G_{m,r}^*\hat{M} + G_{q,r}^*\hat{Q}\right)\hat{A}_r^{\dagger},\left(\frac{G_{m,s}}{\Omega_{a,s} - \Omega_m}\hat{M}^{\dagger} + G_{q,s}\chi_s\left(\hat{Q}^{\dagger}\hat{Q}\right)\hat{Q}^{\dagger}\right)\hat{A}_s\right] + h . c . 
\end{equation} 
For $\hat{f},\hat{g}$ commuting with all photon operators, 
\begin{equation}
	 \left[\hat{f}\hat{A}_r^{\dagger},\hat{g}\hat{A}_s\right] =  - \hat{f}\hat{g}\delta_{rs} + \left[\hat{f},\hat{g}\right]\hat{A}_r^{\dagger}\hat{A}_s . 
\end{equation} 
This leads to 
\begin{eqnarray}
	 \fl \frac{1}{2\hbar}\left[\Ham{int},\hat{S}\right] =  - \sum_r\left(G_{m,r}^*\hat{M} + G_{q,r}^*\hat{Q}\right)\left(\frac{G_{m,r}}{\Omega_{a,r} - \Omega_m}\hat{M}^{\dagger} + G_{q,r}\chi_r\left(\hat{Q}^{\dagger}\hat{Q}\right)\hat{Q}^{\dagger}\right) \nonumber \\ +\delta\Omega_{a,r}\left[\hat{Q}^{\dagger}\hat{Q}\right]\hat{A}_r^{\dagger}\hat{A}_r + h . c . 
\end{eqnarray} 
where 
\begin{equation}
	 \delta\Omega_{a,r}\left[x\right] = \frac{\left|G_{m,r}\right|^2}{\Omega_{a,r} - \Omega_m} + \left|G_{q,r}\right|^2\left(\left(x + 1\right)\chi_r\left(x + 1\right) - x\chi_r\left(x\right)\right) . 
\end{equation} 
This procedure yields the rotated Hamiltonian. To return back to the original one, we define the dressed operators: $\hat{m} = e^{\hat{S}}\hat{M}e^{- \hat{S}}$, $\hat{q} = e^{\hat{S}}\hat{Q}e^{- \hat{S}}$, and $\hat{a}_r = e^{\hat{S}}\hat{A}_re^{- \hat{S}}$. Up to linear order in $G$, the dressed magnon operator is 
\begin{equation}
	 \hat{m} \approx \hat{M} - \sum\frac{G_{m,r}}{\Omega_{a,r} - \Omega_m}\hat{A}_r + O\left(G^2\right), 
\end{equation} 
and the dressed transmon operator is 
\begin{equation}
	 \hat{q} \approx \hat{Q} - \sum G_{q,r}\hat{A}_r\left(\chi_r\left[\hat{Q}\hat{Q}^{\dagger}\right] - \chi_r\left[\hat{Q}^{\dagger}\hat{Q}\right]\right) + O\left(G^2\right) . 
\end{equation} 
This gives the final Hamiltonian 
\begin{equation}
	 \frac{\hat{H}}{\hbar} = \omega_m\hat{m}^{\dagger}\hat{m} + \left(\omega_q + \hat{\alpha}\left[\hat{q}^{\dagger}\hat{q}\right]\right)\hat{q}^{\dagger}\hat{q} + \hat{m}\hat{q}^{\dagger}\hat{g}\left[\hat{q}^{\dagger}\hat{q}\right] + \hat{g}\left[\hat{q}^{\dagger}\hat{q}\right]\hat{m}^{\dagger}\hat{q} + \frac{\Ham{mw,eff}}{\hbar} . 
\end{equation} 
Here, $\Ham{mw,eff}$ is the effective MW Hamiltonian that is not relevant for our discussion. The renormalized magnon's frequency is Stark shifted by the MWs: 
\begin{equation}
	 \omega_m = \Omega_m - \sum_r\frac{\left|G_{m,r}\right|^2}{\Omega_{a,r} - \Omega_m} . 
\end{equation} 
Similarly, the transmon's frequency is 
\begin{equation}
	 \omega_q = \Omega_q - \sum_r\frac{\left|G_{q,r}\right|^2}{\Omega_{a,r} - \Omega_q} . 
\end{equation} 
The Stark shift of higher transmon levels renormalizes the anharmonicity to 
\begin{equation}
	 \hat{\alpha}\left[\hat{q}^{\dagger}\hat{q}\right] = \frac{A}{2}\left(\hat{q}^{\dagger}\hat{q} - 1\right) - \sum_r\left|G_{q,r}\right|^2\left[\frac{1}{\Omega_{a,r} - \Omega_q - A\left(\hat{q}^{\dagger}\hat{q} - 1\right)} - \frac{1}{\Omega_{a,r} - \Omega_q}\right] . 
\end{equation} 
Finally, the coupling of each transmon transition to the magnons is given by 
\begin{equation}
	 \hat{g}\left[\hat{q}^{\dagger}\hat{q}\right] =  - \frac{1}{2}\sum_rG_{m,r}^*G_{q,r}\left(\frac{1}{\Omega_{a,r} - \Omega_q - A\hat{q}^{\dagger}\hat{q}} + \frac{1}{\Omega_{a,r} - \Omega_m}\right) . 
\end{equation}

\section{Transfer of Quanta\label{app:Transfer}} 

In this appendix, we show how to obtain the periods required for transferring quanta from transmon to magnons. In what follows, $\ket{m,q}$ denotes a state with $m$ magnons and transmon at $q$th level. Given a state $\ket{\Psi}$ having a maximum of $j$ excitations, i.e. $\inn{m,q}{\Psi} = 0$ for $m + q > j$ and a tunable Hamiltonian 
\begin{equation}
	 \fl \frac{\hat{H}(t)}{\hbar} = \omega_m\hat{m}^{\dagger}\hat{m} + \left(\omega_q + \Delta(t)\right)\hat{q}^{\dagger}\hat{q} - \sum_{n = 2}^{\infty}\alpha_n\ket{n}\bra{n} + \sum_{n = 0}^{\infty}g_n\hat{m}^{\dagger}\ket{n}\bra{n + 1} + g_n^*\hat{m}\ket{n + 1}\bra{n}, 
\end{equation} 
we want to find $\{\Delta(t),t_{{\rm tot}}\}$ s.t. $\Op{j,0}{\hat{{\cal T}}^{\dagger}}{\Psi} = 0$, where the time-evolution operator is

\begin{equation}
	 \hat{{\cal T}} = \Texp{- \frac{i}{\hbar}\int_0^{t_{{\rm tot}}}\hat{H}(\tau)d\tau}, 
\end{equation} 
with ${\rm Texp}$ being the time-ordered exponential operator. Here, $\alpha_n\equiv\Op{n}{\hat{\alpha}}{n}$ and $g_n\equiv \Op{n}{\hat{g}}{n}$ as derived in the previous section. 

First, consider the ansatz $\Delta(t) = \Delta_0\left[u(t - t_{{\rm on}}) - u(t)\right]$ with $\Delta_0 = \omega_m - \omega_q$, s.t. the magnons and transmons are resonant for a time-period $t_{{\rm on}}$, effectively switching on the coupling. Adding an off-period of $t_{{\rm off}}$ afterwards, we get $t_{{\rm tot}} = t_{{\rm on}} + t_{{\rm off}}$ and we want to find $\{t_{{\rm on}},t_{{\rm off}}\}$ to satisfy the above condition. We can derive the time-evolution unitary explicitly using $[\hat{H}(t),\hat{N}] = 0$ where the total number of excitations are given by 
\begin{equation}
	 \hat{N} = \hat{m}^{\dagger}\hat{m} + \hat{q}^{\dagger}\hat{q} . 
\end{equation} 
Thus, we divide the wave-function as 
\begin{equation}
	 \ket{\Psi(t)} = \sum_{N = 0}^{\infty}\sum_{s = 0}^N\xi_{Ns}(t)\ket{N - s,s} . 
\end{equation} 
During the off-time, we have the equations of motion, [using the Schr\"{o}dinger equation $\ket{\dot{\Psi}} =  - i\hat{H}\ket{\Psi}$]. 
\begin{equation}
	\fl \dot{\xi}_{Ns} =  - i\left(\omega_m\left(N - s\right) + \omega_qs - \alpha_s\right)\xi_{Ns} - i\left[g_s\sqrt{N - s}\ \xi_{N,s + 1} + g_{s - 1}^*\sqrt{N - s + 1}\ \xi_{N,s - 1}\right] . 
\end{equation} 
This is a set of linear differential equations which can be solved by diagonalizing the dynamical matrix. For the case $N = j$, we can write 
\begin{equation}
	 \bi{\xi}_j(t_{{\rm on}} + t_{{\rm off}}) = V_{{\rm off}}e^{- i\Lambda_{{\rm off}}t_{{\rm off}}}V_{{\rm off}}^{\dagger}\bi{\xi}_j(t_{{\rm on}}), 
\end{equation} 
where the vector $\bi{\xi}_j = \left(\xi_{j0}\ \xi_{j1}\ \dots\ \xi_{jj}\right)^T$, and the dynamical matrix is diagonalized by $\{- i\Lambda_{{\rm off}},V_{{\rm off}}\}$. Similar analysis with $\omega_q\rightarrow\omega_m$ works for the on-time, 
\begin{equation}
	 \bi{\xi}_j\left(t_{{\rm on}} + t_{{\rm off}}\right) = V_{{\rm off}}e^{- i\Lambda_{{\rm off}}t_{{\rm off}}}V_{{\rm off}}^{\dagger}V_{{\rm on}}e^{- i\Lambda_{{\rm on}}t_{{\rm on}}}V_{{\rm on}}^{\dagger}\bi{\xi}_j(0) . 
\end{equation} 
We want to ensure $\Op{\Psi}{\hat{{\cal T}}}{j,0} = 0$. We find $\hat{{\cal T}}\ket{j,0}$ by putting $\bi{\xi}_j(0) = \left(1\ 0\ \dots\ 0\right)^T\equiv\bi{u}_j$ and then, the condition $\Op{\Psi}{\hat{{\cal T}}}{j,0} = 0$ becomes 
\begin{equation}
	 \bi{\Psi}_j^{\dagger}\left(V_{{\rm off}}e^{- i\Lambda_{{\rm off}}t_{{\rm off}}}V_{{\rm off}}^{\dagger}V_{{\rm on}}e^{- i\Lambda_{{\rm on}}t_{{\rm on}}}V_{{\rm on}}^{\dagger}\right)\bi{u}_j = 0,\label{eq:to_opt} 
\end{equation} 
where the vector $\bi{\Psi}_j = \left(\inn{j,0}{\Psi}\ \inn{j - 1,1}{\Psi}\ \dots\ \inn{0,j}{\Psi}\right)^T$. This can be numerically solved for $\{t_{{\rm on}},t_{{\rm off}}\}$. 

In the special case of a two-level approximation (ignoring higher levels) and for sufficiently large detuning $g_0\ll\omega_m - \omega_q$, the above condition reduces to 
\begin{equation}
	 \cos\left(g_0t_{{\rm on}}^{(0)}\sqrt{j}\right)\inn{j,0}{\Psi} + ie^{- i\Delta_0t_{{\rm off}}^{(0)}}\sin\left(g_0t_{{\rm on}}^{(0)}\sqrt{j}\right)\inn{j - 1,1}{\Psi} = 0 . 
\end{equation} 
This gives 
\begin{eqnarray}
	 g_0\sqrt{j}t_{{\rm on}}^{(0)} &= \tan^{- 1}\left|\frac{\inn{j,0}{\Psi}}{\inn{j - 1,1}{\Psi}}\right|,\label{eq:app:ton:TLS}\\
	 \Delta_0t_{{\rm off}}^{(0)} &={\rm arg}\left[\frac{\inn{j,0}{\Psi}}{\inn{j - 1,1}{\Psi}}\right] + \frac{\pi}{2} . \label{eq:app:toff:TLS} 
\end{eqnarray}

In some special cases, Eq. (\ref{eq:to_opt}) will have no solution. Then, we try the ansatz with a pulse of width $t_{{\rm on}}^{(0)}$, followed by an off-time $t_{{\rm off}}^{(0)}$ and a second pulse of an undetermined time $t_{{\rm on}}$, 
\begin{equation}
	 \Delta(t) = \Delta_0\left[u\left(t - t_{{\rm on}} - t_{{\rm tot}}^{(0)}\right) - u\left(t - t_{{\rm tot}}^{(0)}\right) + u\left(t - t_{{\rm on}}^{(0)}\right) - u\left(t\right)\right], 
\end{equation} 
where $t_{{\rm tot}}^{(0)} = t_{{\rm on}}^{(0)} + t_{{\rm off}}^{(0)}$. After another off-period of time $t_{{\rm off}}$, the optimization condition becomes (following the same analysis as before) 
\begin{equation}
	 \fl \bi{\Psi}_j^{\dagger}\left(V_{{\rm off}}e^{- i\Lambda_{{\rm off}}t_{{\rm off}}}V_{{\rm off}}^{\dagger}V_{{\rm on}}e^{- i\Lambda_{{\rm on}}t_{{\rm on}}}V_{{\rm on}}^{\dagger}\right)\left(V_{{\rm off}}e^{- i\Lambda_{{\rm off}}t_{{\rm off}}^{(0)}}V_{{\rm off}}^{\dagger}V_{{\rm on}}e^{- i\Lambda_{{\rm on}}t_{{\rm on}}^{(0)}}V_{{\rm on}}^{\dagger}\right)\bi{u}_j = 0, 
\end{equation} 
which can be solved for $\{t_{{\rm on}},t_{{\rm off}}\}$. This process can be repeated indefinitely, however, in practice, we did not observe the need to go beyond a second iteration. 

\section{Transmon Gates\label{app:Gates}} 

We now describe how the transmon gates can be implemented in our scheme. In what follows, $\ket{m,q}$ denotes a state with $m$ magnons and transmon at $q$th level. Given an initial state $\ket{\Psi}$ with a maximum of $j$ magnons and $(j + 1)$ excitations, i.e. $\inn{m,q}{\Psi} = 0$ for $m + q> j + 1$ and $\inn{j + 1,0}{\Psi} = 0$, and a tunable Hamiltonian 
\begin{equation}
	 \frac{\hat{H}(t)}{\hbar} = \omega_m\hat{m}^{\dagger}\hat{m} + \omega_q\hat{q}^{\dagger}\hat{q} - \sum_{n = 2}^{\infty}\alpha_n\ket{n}\bra{n} + \tilde{\varepsilon}(t)\hat{q}^{\dagger} + \tilde{\varepsilon}^*(t)\hat{q}, 
\end{equation} 
we want to find $\tilde{\varepsilon}(t)$ s.t. $\Op{j,1}{\hat{{\cal Q}}^{\dagger}}{\Psi} = 0$, where the time-evolution operator 
\begin{equation}
	 \hat{{\cal Q}} = \Texp{- \frac{i}{\hbar}\int_0^{t_g}\hat{H}(\tau)d\tau}, 
\end{equation} 
with ${\rm Texp}$ being the time ordered exponential operator and $t_g$ being a pre-determined gate time. 

Note that we have ignored the coupling between magnons and transmons assuming large enough detuning $\omega_m - \omega_q$. Then, the evolution is given by 
\begin{equation}
	 \hat{{\cal Q}}(t) = e^{- i\omega_mt\hat{m}^{\dagger}\hat{m}}\otimes\hat{{\cal U}}_g(t) . 
\end{equation} 
Here, 
\begin{equation}
	 \frac{d}{dt}\hat{{\cal U}_g} =  - i\left(\omega_q\hat{q}^{\dagger}\hat{q} - \sum_{n = 2}^{\infty}\alpha_n\ket{n}\bra{n} + \tilde{\varepsilon}(t)\hat{q}^{\dagger} + \tilde{\varepsilon}^*(t)\hat{q}\right)\hat{{\cal U}}_g . 
\end{equation} 
An ideal gate between $\ket{0}$ and $\ket{1}$, with angular parameters $\{\theta,\phi\}$ is 
\begin{equation}
	\fl \hat{{\cal U}}_{{\rm ideal}} = e^{i\Phi_{{\rm global}}}\left[\cos\theta\left(\ket{0}\bra{0} + \ket{1}\bra{1}\right) + i\sin\theta\left(e^{i\phi}\ket{1}\bra{0} + e^{- i\phi}\ket{0}\bra{1}\right) + \sum_{n = 2}^{\infty}\ket{n}\bra{n}\right], 
\end{equation} 
barring an unimportant global phase $\Phi_{{\rm global}}$. The parameters for this gate can be found by ensuring $\Op{j,1}{\hat{{\cal Q}}_{{\rm ideal}}^{\dagger}}{\Psi} = 0$, where $\hat{{\cal Q}}_{{\rm ideal}} = e^{- i\omega_mt_g\hat{m}^{\dagger}\hat{m}}\otimes\hat{{\cal U}}_{{\rm ideal}}$ giving 
\begin{equation}
	 \theta_{{\rm ideal}} = \tan^{- 1}\left|\frac{\inn{j,1}{\Psi}}{\inn{j,0}{\Psi}}\right|\label{eq:theta_ideal} 
\end{equation} 
and 
\begin{equation}
	 \phi_{{\rm ideal}} =  - \frac{\pi}{2} + \text{arg}\left[\frac{\inn{j,1}{\Psi}}{\inn{j,0}{\Psi}}\right] . \label{eq:phi_ideal} 
\end{equation} 
To implement this, we try the ansatz $\tilde{\varepsilon}(t) =  - \varepsilon(t)e^{- i\omega_qt + i\phi_{{\rm app}}}$ with real $\{\varepsilon(t),\phi_{\rm app}\}$ to be found. We separate out the phases by the transformation, 
\begin{equation}
	 \hat{{\cal U}}_g(t) = e^{- i\left(\omega_q\hat{q}^{\dagger}\hat{q} - \sum_{n = 2}^{\infty}\alpha_n\ket{n}\bra{n}\right)t}\hat{{\cal U}}_{{\rm ph}}\hat{{\cal U}}_{{\rm \theta}}(t)\hat{{\cal U}}_{{\rm ph}}^{\dagger},\label{eq:PhaseFactoring} 
\end{equation} 
where 
\begin{equation}
	 \hat{{\cal U}}_{{\rm ph}} = e^{i\phi_{{\rm app}}\hat{q}^{\dagger}\hat{q}} 
\end{equation} 
and $\hat{{\cal U}}_{\theta}$ satisfies 
\begin{equation}
	 \frac{d}{dt}\hat{{\cal U}}_{\theta} = i\varepsilon(t)\left(\sum_{n = 0}^{\infty}\sqrt{n + 1}\ e^{- i\left(\alpha_{n + 1} - \alpha_n\right)t}\ket{n + 1}\bra{n} + h . c . \right)\hat{{\cal U}}_{\theta},\label{eq:UthetaEOM} 
\end{equation} 
where by definition $\alpha_0 = \alpha_1 = 0$. 

First, consider the two-level approximation, where we want to solve 
\begin{equation}
	 \frac{d}{dt}\hat{{\cal U}}_{\theta}^{(0)} = i\varepsilon^{(0)}(t)\left(\ket{0}\bra{1} + \ket{1}\bra{0}\right)\hat{{\cal U}}_{\theta}^{(0)}, 
\end{equation} 
where $\varepsilon^{(0)}(t)$ is the `bare' solution to be chosen. The unitary becomes 
\begin{equation}
	 \fl \hat{{\cal U}}_{\theta}^{(0)}(t) = \cos\Theta(t)\left(\ket{0}\bra{0} + \ket{1}\bra{1}\right) + i\sin\Theta(t)\left(\ket{1}\bra{0} + \ket{0}\bra{1}\right) + \sum_{n = 2}^{\infty}\ket{n}\bra{n}, 
\end{equation} 
where 
\begin{equation}
	 \Theta^{(0)}(t) = \int_0^t\varepsilon^{(0)}(\tau)d\tau . 
\end{equation} 
For such a case, we have 
\begin{eqnarray*}
	\hat{{\cal U}}_{g}^{(0)}(t)=\left[\cos\Theta^{(0)}(t)\left(\ket{0}\bra{0}+e^{-i\omega_{q}t_{g}}\ket{1}\bra{1}\right)+\right.\\
	\left.i\sin\Theta^{(0)}(t)\left(e^{i\phi_{{\rm app}}^{(0)}}e^{-i\omega_{q}t_{g}}\ket{1}\bra{0}+e^{-i\phi_{{\rm app}}^{(0)}}\ket{0}\bra{1}\right)+\sum_{n=2}^{\infty}e^{-i\left(\omega_{q}-\alpha_{n}\right)t}\ket{n}\bra{n}\right].
\end{eqnarray*}

This resembles the ideal gate $\hat{{\cal U}}_{{\rm ideal}}$ except for the extra phases due to the free evolution of the states. For this two-level approximation, we can ensure $\Op{j,1}{\hat{{\cal Q}}^{(0),\dagger}}{\tilde{\Psi}_{j + 1}} = 0$, where $\hat{{\cal Q}}^{(0)} = e^{- i\omega_mt_g\hat{m}^{\dagger}\hat{m}}\otimes\hat{{\cal U}}_g^{(0)}(t_g)$ by choosing 
\begin{equation}
	 \Theta^{(0)}(t_g) = \theta_{{\rm ideal}},\ \phi_{{\rm app}}^{(0)} = \phi_{{\rm ideal}} + \omega_qt_g . \label{eq:TLS_GateParams} 
\end{equation} 
Notice that we can choose any $\varepsilon^{(0)}(t)$ satisfying $\int_0^{t_g}d\tau\varepsilon^{(0)}(\tau) = \theta_{{\rm ideal}}$. For concreteness, we choose a sinusoidal 
\begin{equation}
	 \varepsilon^{(0)}(t) = \frac{\pi\theta_{{\rm ideal}}}{2t_g}\sin\left(\frac{\pi t}{t_g}\right) . 
\end{equation}

\subsection{Including all levels} 

Transition between $\ket{1}$ and $\ket{2}$ are non-resonant by the frequency $\alpha_2$, so the above two-level approximation works when $t_g\gg\alpha_2$ [assuming that the anharmonicity increases with levels, i.e. $\alpha_2<\alpha_3<\alpha_4<\dots$]. For a general $t_g$, we modify the pulse $\varepsilon(t) = \varepsilon^{(0)}(t) + \varepsilon^{(1)}(t)$ and define the error as $\hat{{\cal U}}_{\theta} = \hat{{\cal U}}_{\theta}^{(0)}\hat{{\cal U}}_{{\rm err}}$. The error satisfies 
\begin{equation}
	 \frac{d\hat{{\cal U}}_{{\rm err}}}{dt} = i\delta\hat{H}\ \hat{{\cal U}}_{{\rm err}}, 
\end{equation} 
where the error Hamiltonian is 
\begin{equation}
	 \fl \delta\hat{H} = \hat{{\cal U}}_{\theta}^{(0)\dagger}\left(\varepsilon(t)\sum_{n = 0}^{\infty}\sqrt{n + 1}\ e^{- i\left(\alpha_{n + 1} - \alpha_n\right)t}\ket{n + 1}\bra{n} - \varepsilon^{(0)}(t)\left(\ket{0}\bra{1} + \ket{1}\bra{0}\right) + h . c . \right)\hat{{\cal U}}_{\theta}^{(0)} . 
\end{equation} 
This can be divided as 
\begin{equation}
	 \delta\hat{H} = \left[\varepsilon(t)\left(\delta\hat{h}_{12} + \delta\hat{h}_{{\rm rest}}\right) + \left(\varepsilon(t) - \varepsilon^{(0)}(t)\right)\delta\hat{h}_{01}\right] 
\end{equation} 
with the reduced Hamiltonians defined as 
\begin{eqnarray*}
	\delta\hat{h}_{01} & =\ket{0}\bra{1}+\ket{1}\bra{0}\\
	\delta\hat{h}_{12} & =\sqrt{2}\ e^{-i\alpha_{2}t}\left[\cos\Theta(t)\ket{2}\bra{1}+i\sin\Theta(t)\ket{2}\bra{0}\right]+h.c.\\
	\delta\hat{h}_{{\rm rest}} & =\sum_{n=2}^{\infty}\sqrt{n+1}\ e^{-i\left(\alpha_{n+1}-\alpha_{n}\right)t}\ket{n+1}\bra{n}+h.c.
\end{eqnarray*}

We perform a Magnus expansion to find \cite{MagnusCorr17,MagnusCorr21} 
\begin{equation}
	 \hat{{\cal U}}_{{\rm err}}(t) = \exp\left[i\int_0^td\tau\delta\hat{H}(\tau) - \frac{1}{2}\int_0^td\tau_1\int_0^{\tau_1}d\tau_2\left[\delta\hat{H}(\tau_1),\delta\hat{H}(\tau_2)\right] + \dots\right], 
\end{equation} 
so, to ensure $\hat{{\cal U}}_{{\rm err}}(t_g) \approx \hat{I}$ up to first order, we require $\int_0^{t_g}d\tau\delta\hat{H}(\tau) = \hat{0}$ which translates to 
\begin{eqnarray}
	 \int_0^{t_g}\varepsilon(\tau)d\tau &= \theta_{{\rm ideal}},\label{eq:app:Conds1}\\
	 \int_0^{t_g}\varepsilon(\tau)e^{iA_2\tau\pm i\Theta(t)}d\tau &= 0,\label{eq:app:Conds2}\\ 
	 \int_0^{t_g}\varepsilon(\tau)e^{i\left(A_3 - A_2\right)\tau}d\tau &= 0 . \label{eq:app:Conds3} 
\end{eqnarray} 
We used $\int_0^{t_g}\varepsilon^{(0)}d\tau = \theta_{{\rm ideal}}$. The last condition ensures that there are no $\ket{2}\leftrightarrow\ket{3}$ transitions, and we ignored all higher transitions as they are approximately unoccupied throughout the protocol. 

\subsection{Satisfying the conditions} 

The conditions allow for a lot of freedom in choosing $\varepsilon(t)$. Here, we take an ansatz of a sum of sinusoids, ensuring that the pulse starts and ends at $0$, 
\begin{equation}
	 \varepsilon(t) = \sum_{\nu}\varepsilon_{\nu}\sin\left(\frac{\nu\pi t}{t_g}\right), 
\end{equation} 
where the set of indices $\nu$ can be freely chosen. Eq. (\ref{eq:app:Conds1}) reads 
\begin{equation}
	 \sum_{\nu\:{\rm odd}}\frac{2\varepsilon_{\nu}t_g}{\pi\nu} = \theta_{{\rm ideal}} . \label{eq:lin_cond:theta} 
\end{equation} 
Eq. (\ref{eq:app:Conds3}) is 
\begin{equation}
	 \sum_{\nu}\frac{\pi\nu\varepsilon_{\mu}t_g}{\left(\alpha_3 - \alpha_2\right)^2t_g^2 - \pi^2\nu^2}\left(1 - \left( - 1\right)^{\nu}e^{i\left(\alpha_3 - \alpha_2\right)t_g}\right) = 0 . 
\end{equation} 
Eq. (\ref{eq:app:Conds2}) is 
\begin{equation}
	 e^{\pm i\theta/2}\sum_{\nu}\frac{\varepsilon_{\mu}t_g}{\pi}\int_0^{\pi}du\sin\left(\nu u\right)\exp\left[i\frac{\alpha_2t_g}{\pi}u\mp\frac{i\theta}{2}\cos u\right] = 0 . 
\end{equation} 
To simplify the integration, we define a real integral 
\begin{equation}
	 G_{\zeta}(\xi) = \int_0^{\pi/2}du\cos\left[\zeta u - \xi\sin u\right] 
\end{equation} 
giving 
\begin{equation}
	 \int_0^{\pi}du\exp\left[i\zeta u + i\xi\cos u\right] = 2i^{\zeta}G_{\zeta}(\xi) . 
\end{equation} 
For $\zeta = A_2t_g/\pi$, this gives 
\begin{eqnarray}
	0 &= \sum_{\nu\:{\rm odd}}i^{\nu - 1}\frac{\varepsilon_{\nu}t_g}{\pi}\left[G_{\zeta + \nu}\left(\frac{\pm\theta}{2}\right) + G_{\zeta - \nu}\left(\frac{\pm\theta}{2}\right)\right] \\ &= \sum_{\nu\:{\rm even}}i^{\nu}\frac{\varepsilon_{\nu}t_g}{\pi}\left[G_{\zeta + \nu}\left(\frac{\pm\theta}{2}\right) - G_{\zeta - \nu}\left(\frac{\pm\theta}{2}\right)\right] . 
\end{eqnarray} 
If we choose the gate time s.t. $\left(A_3 - A_2\right)t_g = 2N\pi$ for some integer $N$, we find that the odd and even frequency components can be separated, and in particular, we need only the odd components to satisfy all the above conditions. 

For simulation purposes, we choose the specific case of $A_n = An(n - 1)/2$, the bare transmon, and choose $At_g = N\pi$ for some natural number $N$ giving (for some arbitrary set of indices $H$), 
\begin{eqnarray}
	\varepsilon(t) & =\sum_{\mu\in H}\varepsilon_{\mu}\sin\left(\frac{\left(2\mu+1\right)\pi t}{t_{g}}\right)\\
	\frac{\pi\theta_{{\rm ideal}}}{2t_{g}} & =\sum_{\mu\in H}\frac{\varepsilon_{\mu}}{2\mu+1}\\
	0 & =\sum_{\mu\in H}\frac{\left(2\mu+1\right)\varepsilon_{\mu}}{4N^{2}-\left(2\mu+1\right)^{2}}\\
	0 & =\sum_{\mu\in H}\left(-1\right)^{\mu}\varepsilon_{\mu}\frac{G_{N+2\mu+1}(\pm\theta/2)+G_{N-2\mu-1}\left(\pm\theta/2\right)}{2}
\end{eqnarray}

This is a set of linear equations which can be solved to give $\varepsilon_{\mu}$. To find an expression for $G_n(\theta)$, we use the fact that it satisfies the same recursion relation as that of Bessel functions 
\begin{equation}
	 2G_n'(\xi) = G_{n - 1}(\xi) - G_{n + 1}(\xi) 
\end{equation} 
with the intial conditions, 
\begin{equation}
	 G_0(\xi) - \frac{\pi}{2}J_0(\xi) = 0,\ G_1(\xi) - \frac{\pi}{2}J_1(\xi) = \frac{\sin\xi}{\xi} . 
\end{equation} 
Then, an expression for $G_n(\xi) - \pi J_n(\xi)/2$ can be found recursively. 

\subsection{Phase\label{subapp:Phase_Gate}} 

Given $\varepsilon(t)$, we find $\phi_{{\rm app}}$ to minimize $\left|\Op{j,1}{\hat{{\cal Q}}^{\dagger}}{\Psi}\right|$. We have $\hat{{\cal Q}}(t) = e^{- i\omega_mt\hat{m}^{\dagger}\hat{m}}\otimes\hat{{\cal U}}_g(t)$ where $\hat{{\cal U}}_g(t)$ can be decomposed as in Eq. (\ref{eq:PhaseFactoring})]. Then,
\begin{equation}
	 \hat{{\cal U}_g}(t_g)\ket{1} = e^{- i\phi_{{\rm app}}}e^{- i\left(\omega_q\hat{q}^{\dagger}\hat{q} - \sum_{n = 2}^{\infty}\alpha_n\ket{n}\bra{n}\right)t_g}\hat{{\cal U}}_{{\rm ph}}\hat{{\cal U}}_{{\rm \theta}}(t_g)\ket{1} . 
\end{equation} 
As the leakage into higher levels is suppressed via an appropriate choice of $\varepsilon(t)$, we expect the 2x2 sub-block of $\hat{{\cal U}}_g(t_g)$ for the subspace spanned by $\{\ket{0},\ket{1}\}$ to be approximately unitary. For concreteness, we find the unitary matrix closest to this sub-block, say $\hat{\cal U}_{g,01}$, in 2-norm via $\hat{\cal U}_{g,01}^{{\rm unit}} = VW^{\dagger}$ using the singular value decomposition $\hat{\cal U}_{g,01} = V\Lambda W^{\dagger}$. This gives 
\begin{equation}
	\hat{\cal U}_{g,01}^{{\rm unit}}\ket{1} = e^{i\Phi_G}\left(i\sin\tilde{\theta}e^{i\delta\phi}\ket{0} + \cos\tilde{\theta}\ket{1}\right), 
\end{equation} 
for some constants $\{\Phi_G,\tilde{\theta},\delta\phi\}$. Then, 
\begin{equation}
	 \fl \Op{\Psi}{\hat{\cal U}_g(t_g)}{j,1} \approx e^{- ij\omega_mt - i\phi_{{\rm app}} + i\Phi_G}\left(i\sin\tilde{\theta}e^{i\delta\phi}\inn{\Psi}{j,0} + \cos\tilde{\theta}e^{i\phi_{{\rm app}} - i\omega_qt_g}\inn{\Psi}{j,1}\right), 
\end{equation} 
giving the amplitude 
\begin{equation}
	 \fl \left|\Op{\Psi}{\hat{\cal U}_g(t_g)}{j,1}\right| \approx \sqrt{\left|\inn{\Psi}{j,0}\right|^2 + \left|\inn{\Psi}{j,1}\right|^2}\left|\cos\theta_{{\rm ideal}}\sin\tilde{\theta} - \cos\tilde{\theta}\sin\theta_{{\rm ideal}}e^{i\phi_{{\rm app}} - i\phi_{{\rm app}}^{(0)} - i\delta\phi}\right| . 
\end{equation} 
 This is minimized at 
\begin{equation}
	 \phi_{{\rm app}} = \phi_{{\rm app}}^{(0)} + \delta\phi . 
\end{equation} 
 
\subsection{Gates between higher levels} 

Typically, there would be leakage from magnons into higher levels of transmon, which needs to be corrected via applying gates between the higher levels. Consider the case when we want to apply a gate between $\ket{n}$ and $\ket{n + 1}$ with angular parameters $\{\theta,\phi\}$. We try for a pulse shape 
\begin{equation}
	 \tilde{\varepsilon}(t) =  - \frac{\varepsilon(t)}{\sqrt{n}}e^{i\phi_{{\rm app}}}e^{- i\left(n\omega_m - \alpha_n\right)t} . 
\end{equation} 
Again, we have the `bare' pulse $\varepsilon^{(0)}$ which we take as half-sines. To remove the errors at first order in the Magnus expansion, we find that $\varepsilon(t)$ here satisfies the same set of conditions Eqs. (\ref{eq:app:Conds1})-(\ref{eq:app:Conds3}). Thus, the same calculations as above apply.  

\section*{References}

\bibliographystyle{unsrt}
\bibliography{References} 

\end{document}